\pdfoutput=1
\documentclass[aps,prd,amsmath,floats,floatfix,onecolumn,superscriptaddress,nofootinbib,showpacs]{revtex4-2}

\usepackage[T1]{fontenc}
\usepackage[utf8]{inputenc}
\usepackage{lmodern}
\usepackage{lipsum}
\usepackage{physics}

\usepackage[dvipsnames, usenames]{xcolor}
\definecolor{linkcolor}{rgb}{0.0,0.3,0.5}
\usepackage[hypertexnames=false, unicode, colorlinks=true, linkcolor=linkcolor,
citecolor=linkcolor, filecolor=linkcolor,urlcolor=linkcolor,
pdfusetitle]{hyperref}

\usepackage{import}

\usepackage[all]{hypcap}
\usepackage{graphicx}
\usepackage{xspace}
\usepackage{amssymb}
\usepackage[normalem]{ulem} 
\usepackage{bm} 
\usepackage{appendix}
\usepackage[small]{caption}
\usepackage{subcaption}

\usepackage{microtype}

\usepackage[english]{babel}
\usepackage{blindtext}

\graphicspath{%
  {figs/}%
}

\usepackage{tikz}
\usetikzlibrary{shapes,snakes}
\usetikzlibrary{arrows,intersections,patterns,decorations.markings,shapes.geometric}
\usetikzlibrary{calc,fadings,decorations.pathreplacing,positioning}
\usetikzlibrary{decorations.pathmorphing}
\tikzset{snake it/.style={decorate, decoration=snake}}

\usepackage{pgfplots}
\usepgfplotslibrary{colormaps}
\usetikzlibrary{intersections,patterns,pgfplots.fillbetween}

\tikzset{->-/.style={decoration={
  markings,
  mark=at position .5 with {\arrow{>}}},postaction={decorate}}
}
\tikzset{-<-/.style={decoration={
  markings,
  mark=at position .5 with {\arrow{<}}},postaction={decorate}}
}

\tikzset{%
  >=latex, 
  inner sep=0pt,%
  outer sep=2pt,%
  mark coordinate/.style={inner sep=0pt,outer sep=0pt,minimum size=3pt,
    fill=black,circle}%
}

\DeclareMathAlphabet{\mathpzc}{OT1}{pzc}{m}{it}

\renewcommand{\vec}[1] {\bm{#1}}
\newcommand{\vhat}[1]{\vec{\hat{#1}}}

\newcommand{\dchi}{\left.\frac{d\chi}{d\delta\chi}\right|_{\delta\chi=0}}

\newcommand\cevarr[1]{\vec{\overleftarrow{#1}}}

\newcommand{\ClUT}{\mathbb{C}}

\newcommand{\fnl}{f_{\rm NL}}

\newcommand{\pr}{^{\prime}}
\def\VEV#1{\left\langle #1 \right\rangle}

\definecolor{darkgreen}{RGB}{1,150,24}

\definecolor{darkred}{RGB}{175,0,0}
\definecolor{darkblue}{RGB}{14,0,185}
\definecolor{salmon}{RGB}{255,160,105}

\definecolor{darkteal}{RGB}{0,121,150}

\begin{document}

\rightline{\scriptsize RBI-ThPhys-2025-11}

\title{Observed unequal-time power spectrum}

\author{Francesco Spezzati}
\affiliation{Dipartimento di Fisica Galileo Galilei, Universit\` a di Padova, I-35131 Padova, Italy}
\affiliation{INFN Sezione di Padova, I-35131 Padova, Italy}

\author{Eleonora Vanzan}
\affiliation{Dipartimento di Fisica Galileo Galilei, Universit\` a di Padova, I-35131 Padova, Italy}
\affiliation{INFN Sezione di Padova, I-35131 Padova, Italy}
\affiliation{School of Physics and Astronomy, Tel-Aviv University, Tel-Aviv 69978, Israel}

\author{Alvise Raccanelli}
\affiliation{Dipartimento di Fisica Galileo Galilei, Universit\` a di Padova, I-35131 Padova, Italy}
\affiliation{INFN Sezione di Padova, I-35131 Padova, Italy}
\affiliation{INAF-Osservatorio Astronomico di Padova, Italy}

\author{Zvonimir Vlah}
\affiliation{Ru\dj er Bo\v{s}kovi\'c Institute, Bijeni\v{c}ka cesta 54, 10000 Zagreb, Croatia}
\affiliation{Kavli Institute for Cosmology, University of Cambridge, Cambridge CB3 0HA, UK}
\affiliation{Department of Applied Mathematics and Theoretical Physics, University of Cambridge, Cambridge CB3 0WA, UK }

\author{Daniele Bertacca}
\affiliation{Dipartimento di Fisica Galileo Galilei, Universit\` a di Padova, I-35131 Padova, Italy}
\affiliation{INFN Sezione di Padova, I-35131 Padova, Italy}
\affiliation{INAF-Osservatorio Astronomico di Padova, Italy}

\begin{abstract}
The next generation of galaxy surveys will provide highly precise measurements of galaxy clustering, therefore requiring a corresponding accuracy. Current approaches, which rely on approximations and idealized assumptions, may fall short in capturing the level of detail required for high-precision observations. In order to increase the modeling accuracy,
recently, unequal-time contributions to the galaxy power spectrum have been introduced in order to include the effects of radial correlations. We present a generalization of the formalism for the observed unequal-time power spectrum, that includes Doppler and local general relativistic corrections, plus local primordial non-Gaussianity.
We find that unequal time corrections can potentially mimic an effective $f_{\mathrm{NL}}$ of order unity.
We provide a first assessment of the significance of unequal-time corrections for future galaxy clustering experiments, estimating a Signal-to-Noise-Ratio of~$\sim3$ for Stage IV-{\it like} surveys.
\end{abstract} 

\maketitle

\section{Introduction}
\label{sec:intro}

The study of galaxy clustering is about to enter its most fruitful era. New ground-based and space-borne instruments, such as Euclid~\cite{Amendola2016}, DESI~\cite{DESI2016}, Rubin~\cite{LSSTDarkEnergyScience2012}, Roman~\cite{Spergel2015}, SPHEREx~\cite{SPHEREx2014}, SKAO~\cite{SKA2018}, promise to provide data of unparalleled precision. Yet, in order to obtain unbiased constraints on cosmological parameters, it is critical to have a matching degree of accuracy on the theoretical modeling side.

In the past, small survey volumes allowed to work within the flat-sky and distant-observer approximations, thus maintaining a relatively simple description of the statistical observables.
Now, in order to capture in an unbiased way the full information that next generation datasets will deliver, we need to reconsider this and might need to drop these approximations.
In order to properly describe the observed galaxy clustering on the light-cone, one need to account for geometry and observational effects, and a general relativistic treatment of cosmological perturbations~\cite{Bertacca2012, Bertacca2019,Jeong2011,Jeong:2014ufa,Castorina:2021xzs,Schmidt:2012ne,Mitsou:2019ocs}.
Failing to do so might cause biasing results (see e.g.,~\cite{Raccanelli2015, Bellomo2020, Bernal2020}). 

At the level of the two-point correlation function (2PCF), the importance of wide-angle and geometrical effects has been investigated, e.g.,~in~\cite{Szalay1997, Szapudi2004, Papai2008, Raccanelli2010, Desjacques2020, Spezzati2024,Yoo:2013zga,Raccanelli2015}.
Projection effects which arise from gravitational potentials (gravitational lensing, integrated Sachs-Wolfe, and time delay effects) have been investigated, e.g.,~in~\cite{Bertacca2012,Raccanelli2015,Grimm2020,Yoo2009,Yoo2010,Bonvin2011,Challinor2011,Jeong2011,Maartens2012,Tansella2017,Yoo2019,Semenzato2024,DiDio:2020jvo,Camera:2014bwa}. 

Recently, a new formalism was developed to account for an important but often overlooked aspect of galaxy correlations: sources being correlated are generally not located at the same redshift, introducing what can be called an “unequal-time” effect. This correction can be incorporated in Fourier space through a Taylor expansion in the separation between tracers, allowing us to capture radial information that would otherwise be lost when galaxies within a redshift bin are approximated as lying at the same mean redshift.
The importance of properly modeling these radial correlations has been demonstrated both in Fourier space and in configuration space. The latter case was investigated in detail in~\cite{Spezzati2024}, which showed that accurate theoretical descriptions require proper treatment of radial modes.

This paper is organized as follows.
Section~\ref{sec:pk} reviews the unequal-time formalism, and introduces General Relativistic and Doppler terms in redshift space in the theoretical model.
Section~\ref{sec:calculation} computes the observed power spectrum, Section~\ref{sec:grterms} works out the details of the unequal-time coefficients, Section~\ref{sec:PnG} includes local-type primordial non-Gaussianity.
Finally,~\ref{sec:relevance} provides a first assessment of the significance of unequal-time corrections for future galaxy surveys.

\section{Unequal-time galaxy power spectrum in redshift space}
\label{sec:pk}

In this section, we briefly review the basics of the unequal-time galaxy power spectrum; we refer the reader to~\cite{RVa,RVb,Gao2023}  for more details.
The ``true'' power spectrum is the ensemble average of two galaxy overdensity fields, in general evaluated at different times,
\begin{equation}
\label{eq:phyper}
    \langle \delta(\vec{k},t) \delta^*(\vec{k}',t') \rangle = (2\pi)^3 \delta^{(3)}_D(\vec{k}-\vec{k}') \mathcal{P}(\vec{k},t,t') \, .
\end{equation}
This object is in principle fully accessible only to a hypothetical meta-observer, i.e.,~one that could observe the full evolution of the overdensity field in 4D (in what we call the Hyperuranion space~\cite{RVa}). 
When we observe the sky, we only have access to information on the light-cone, thus we are limited to observables that can be constructed from measured quantities, e.g.,~angular positions and redshifts.
While a meta-observer can access the full 4D information, for an the observer on the light-cone there is a degeneracy between cosmic time and the radial direction component of each Fourier mode (for a visualization of the setup, see~\cite{RVa, RVb}.
The most straightforward and correct way to maintain all the radial information provided by a survey consists in slicing the galaxy catalog in many thin redshift bins and compute the angular power spectra including all cross-correlations. While correct, this has the cost of having to compute a prohibitively large number of cross-bin correlations, with non-diagonal covariance for cross-correlations. Alternatively, the spherical Fourier-Bessel formalism naturally includes all radial modes, but these again come at extreme computational costs, on top of the problem of having the information spread over series of angular and Fourier modes (see e.g.,~\cite{Heavens:1994iq, 2dFGRS:2004cmo, Yoo:2013tc, Semenzato2024}).

Recently,~\cite{RVa,RVb} proposed a novel way to retain the unequal-time information in the power spectrum, while still maintaining a relatively low computational cost, both in Fourier space and with angular correlations. While the true dynamics of galaxy clustering happens in the 4D Hyperuranion space, where the meta-observer has access to 3 spatial dimensions at every instant in time, observations happen on the past lightcone of an observer.
To build a proper observable, one then needs to connect the two, folding the 4D observable into a 3D one on the lightcone that includes the effect of radial/unequal in time modes.

The procedure consists of two steps: first the galaxy overdensity field is projected on the sky to build the full unequal-time angular power spectrum, then the information on the unequalness-in-time between the two galaxies is folded into the radial Fourier modes; see~\cite{RVa, RVb} for details and pictorial representations.

We start by constructing the unequal-time galaxy angular power spectrum, including projection effects such as the standard Kaiser term, Doppler terms, and general relativistic corrections.
To obtain an accurate picture of the importance of UT effects, it is necessary to include in the modeling all the effects that are expected to impact the large scales, such as Doppler term and general relativistic effects ~\cite{Yoo2010, Bonvin2011, Challinor2011, Bertacca2012, Raccanelli2016, Borzyszkowski2017, Bertacca2019, Elkhashab:2021lsk, Euclid:2024ufa}.
Moreover, we will include the effect of local-type Primordial non-Gaussianity, which induces a long-short mode coupling that affects halo clustering at late time, through a scale-dependent modification of the halo bias which also impacts the large scales~\cite{Matarrese2008, Dalal2007, Slosar2008, Verde2009, Desjacques2010}.
In this paper we focus on redshifts larger than~$z\gtrsim1$, where the global plane parallel approximation assumed so far is valid. In the rest of the paper we will refer to our assumption as ``flat sky'', in the sense that we assume the sky to be flattened, and not curved, as it was done in~\cite{Gao2023}. For a revisited treatment of the flat sky angular power spectrum see also~\cite{Gao:2023tcd}.
Lower redshifts can be analyzed using one of the currently available approaches in angular~\cite{Yoo:2009au,Yoo:2010ni,Bonvin2011,Challinor2011}, configuration~\cite{Bertacca2012, Raccanelli2016, Raccanelli:2013gja, Raccanelli:2013dza}, or spherical~\cite{Heavens:1994iq, 2dFGRS:2004cmo, Yoo:2013tc, Semenzato2024} spaces, or with an extension of this approach that includes sky curvature.

The unequal-time approach promises to provide a useful and more accurate way to analyze galaxy clustering, and as such there will be more progress needed for it to be applied to real data.
A slightly different approach based on field level analyses is presented in~\cite{Steele:2025djd}, where unequal-time corrections are computed also in the case where one projects the fields individually before correlating them, and including cross-bin correlations.
Here we focus on extending the theoretical modeling for the correlator approach.

Note that in what follows we do not account for integrated general relativistic effects such as lensing and time delays.
Including these type of terms in the unequal-time angular power spectrum presented in~\cite{RVa,RVb} is the focus of another ongoing work.


\subsection{Unequal-time angular power spectrum in redshift space}
\label{sec:calculation}

In the synchronous-comoving gauge, the line element reads
\begin{equation}
    ds^2=a(\tau)^2\big\{-d\tau^2+\left[(1-2\mathcal{R})\delta_{ij}+2\partial_i\partial_jE\right]dx^idx^j\big\} \, ,
\end{equation}
where~$\tau$ is the conformal time, $\mathcal{R}$ and~$E$ are scalar perturbations, and~$\mathcal{R}'=0$ in $\Lambda$CDM, where the prime denotes derivative with respect to conformal time.
The observed galaxy overdensity at redshift~$z$ and direction~$\vhat{n}$ is~\cite{Jeong2011,Bertacca2012}
\begin{equation}
\begin{split}
    \delta_g(\vhat{n},z) =& b \delta +\left[ b_e -1 -2Q +\frac{1+z}{H}\frac{dH}{dz} -\frac{2(1-Q)(1+z)}{H\chi} \right] (\partial_{\|}E' +E'') \\
    &-\frac{1+z}{H} \partial_{\|}^2 E' -\frac{2}{\chi}(1-Q) (\chi \mathcal{R} +E') \, ,
\end{split}
\end{equation}
where~$\partial_{\|}=\hat{n}^i\partial_i$, $\chi(z)$ is the comoving distance, $b(z)$ is the linear bias, $$b_e(z) = -(1+z) \frac{\partial\ln N_g(z)}{\partial z}$$ is the evolution bias, and
$$Q(z) = \left.\frac{\partial\ln N_g}{\partial\ln L}\right|_{L=L_{\rm min}}$$
is the magnification bias, with~$N_g$ the number of galaxies with luminosity~$L > L_{\rm min}$ at redshift~$z$. 
The metric perturbations can be related to the matter density contrast in the synchronous gauge~\cite{Jeong2011, Bertacca2012}
In $\Lambda$CDM+GR this leads to 
\begin{equation}
    \delta_g(\vhat{n},z) = \left[ b +f \partial_{\|}^2\nabla^{-2} -\frac{H^2}{(1+z)^2} \frac{3}{2}\Omega_m \mathcal{B} \nabla^{-2} +\frac{f\alpha}{\chi}\partial_{\|}\nabla^{-2} \right] \delta \, ,
\end{equation}
where $f=d\ln D/\ln a$ is the growth rate and the redshift dependent quantities~$\alpha$ and~$\mathcal{B}$ read
\begin{align}
\label{eq:alpha}
    \alpha(b_e,Q,z) &= -\frac{H(z)\chi(z)}{1+z}\qty[ b_e(z) -1 -2 Q(z) +\frac{3}{2} \Omega_m(z) -\frac{2(1-Q(z))(1+z)}{H(z)\chi(z)} ] \, , \\
    \mathcal{B}(b_e,Q,z) &= b_e(z) \left(1-\frac{2f(z)}{3\Omega_m(z)}\right) +1 +\frac{2f}{\Omega_m(z)} +\frac{3}{2}\Omega_m(z) -\frac{2(1-Q(z))(1+z)}{H(z) \chi(z)} -4Q(z) -f(z) \, .
\end{align}

Let~$\vec{\theta}$ be the transverse vector with respect to the line of sight~$\vhat{n}$, so that~$\vec{x}=\chi(\vhat{n}+\vec{\theta})$.
The projected observed overdensity field is
\begin{equation}
    \delta_g(\vec{\theta},z[\bar{\chi}]) = \int d\chi W(\chi,\bar{\chi}) \delta_g(\vec{x}=\chi(\vhat{n}+\vec{\theta}),z[\chi]) \, ,
\end{equation}
where~$W(\chi,\bar{\chi})$ is a window function centered at~$\bar{\chi}$.
For now we are keeping the explicit dependence on the window function, but in the following we will be interested in the case where it is infinitely thin, so that~$\chi=\bar{\chi}$.
In Fourier space,
\begin{equation}
\begin{split}
    \delta_g(\vec{\ell},z[\bar{\chi}])= \int d^2\vec{\theta} e^{-i\vec{\ell}\cdot\vec{\theta}} \int d\chi W(\chi,\bar{\chi})D(z[\chi]) \left[ b +f \partial_{\|}^2\nabla^{-2} -\frac{H^2}{(1+z)^2} \frac{3}{2} \Omega_m \mathcal{B} \nabla^{-2} +\frac{f\alpha}{\chi} \partial_{\|}\nabla^{-2} \right] \delta(\vec{x},z[\chi]) \, .
\end{split}
\end{equation}
The operators acting on the matter overdensity become
\begin{equation}
\begin{split}
     \partial_{\|}^a\nabla^{-2}\delta(\vec{x},z[\chi]) &= \partial_{\|}^a\nabla^{-2}\int \frac{d^3\vec{k}}{(2\pi)^3} e^{i\chi\vec{k}\cdot(\vhat{n}+\vec\theta)} \delta(\vec{k}=k_{\hat{n}}\vhat{n}+\vec{k}_{\perp},z[\chi]) \\
     &= i^{(a+2)} \int \frac{d^3\vec{k}}{(2\pi)^3} e^{i\chi\vec{k}\cdot(\vhat{n}+\vec\theta)} \frac{(\vec{k}\cdot\vhat{n})^a}{k^2} \delta(\vec{k}=k_{\hat{n}}\vhat{n}+\vec{k}_{\perp},z[\chi]) \, ,
\end{split}
\end{equation}
therefore, performing the angular integration, we get
\begin{equation}
\begin{split}
    \delta_g(\vec{\ell},z[\bar{\chi}]) =& \int d\chi W(\chi,\bar{\chi}) \int \frac{dk_{\hat{n}}}{2\pi} \frac{d^2\vec{k}_{\perp}}{(2\pi)^2} e^{i\chi \vec{k}\cdot\vhat{n}} (2\pi)^2 \delta_D^{(2)}(\vec{\ell}-\chi\vec{k}_{\perp}) \\
    & \qquad D(z[\chi]) \left[ b +f \frac{(\vec{k}\cdot\vhat{n})^2}{k^2} +\frac{H^2}{(1+z)^2} \frac{3}{2} \Omega_m \mathcal{B} \frac{1}{k^2} -i\frac{f\alpha}{\chi} \frac{(\vec{k}\cdot\vhat{n})}{k^2} \right] \delta(\vec{k}=k_{\hat{n}}\vhat{n}+\vec{k}_{\perp},z[\chi])  \, ,
\end{split}
\end{equation}
where the Dirac delta relates the transverse Fourier mode to the angular scale, and integrating over~$\vec{k}_{\perp}$ we obtain
\begin{equation}
\label{eq:ell}
    \delta_g(\vec{\ell},z[\bar{\chi}]) = \int \frac{d\chi}{\chi^2} W(\chi,\bar{\chi}) D(z[\chi]) \int \frac{dk_{\hat{n}}}{2\pi} e^{ik_{\hat{n}}\chi} \left[ b +f\mu^2 +\frac{1}{k^2} \frac{H^2}{(1+z)^2} \frac{3}{2} \Omega_m \mathcal{B} -i\frac{f\mu\alpha}{k\chi} \right] \delta(\vec{k}=k_{\hat{n}}\vhat{\vec{n}}+\vec{\ell}/\chi,z[\chi]) \, ,
\end{equation}
where note that~$\mu(k_{\hat{n}},\ell/\chi)$ and~$k(k_{\hat{n}},\ell/\chi)$.
The two-point unequal-time correlator of the projected overdensity field, Equation~\eqref{eq:ell}, for two different dark matter tracers~$A$ and~$B$ at redshifts~$z[\bar{\chi}_1]$ and~$z[\bar{\chi}_2]$ respectively, is 
\begin{equation}
\label{eq:correlator}
\begin{split}
    & \VEV{ \delta_{g,A}(\vec{\ell}_1,z_1[\bar{\chi_1}]) \delta_{g,B}^*(\vec{\ell}_2,z_2[\bar{\chi}_2]) } \\
    &= \int \frac{d\chi_1}{\chi_1^2}\frac{d\chi_2}{\chi_2^2} W(\chi_1,\bar{\chi}_1)  W(\chi_2,\bar{\chi}_2) D(z[\chi_1)]) D(z[\chi_2]) \int \frac{dk_{1,\hat{n}}}{2\pi} \frac{dk_{2,\hat{n}}}{2\pi} e^{ik_{1,\hat{n}}\chi_1} e^{-ik_{2,\hat{n}}\chi_2} \\
    &\qquad \left[b_{1,A}
     +f_1 \mu_1^2 +\frac{\mathcal{A}_{1,A}}{k_1^2}
     -i\frac{f_1\mu_1\alpha_{1,A}}{k_1\chi_1} \right] \left[ b_{2,B} +f _2\mu_2^2 +\frac{\mathcal{A}_{2,B}}{k_2^2} +i\frac{f_2\mu_2\alpha_{2,B}}{k_2\chi_2} \right] \\
     &\qquad \VEV{ \delta(k_{1,\hat{n}}\vhat{n}, \vec{\ell}_1/\chi_1, z_1[\chi_1]) \delta^*(k_{2,\hat{n}}\vhat{n},\vec{\ell}_2/\chi_2,z_2[\chi_2]) } \\
     &= \int \frac{d\chi_1}{\chi_1^2}\frac{d\chi_2}{\chi_2^2} W(\chi_1,\bar{\chi}_1)  W(\chi_2,\bar{\chi}_2) (2\pi)^2 \delta_D^{(2)}\left(\frac{\vec{\ell}_1}{\chi_1}-\frac{\vec{\ell}_2}{\chi_2}\right) D(z[\chi_1)]) D(z[\chi_2]) \\
     &\qquad \int \frac{dk_{\hat{n}}}{2\pi} e^{ik_{\hat{n}}(\chi_1-\chi_2)} \mathcal{P}(k_{\hat{n}}\vhat{n},\vec{k}_{\perp},\chi_1,\chi_2) K_A(k,\mu_1,\chi_1) K_B(k,\mu_2,\chi_2) \, .
\end{split}
\end{equation}
We have defined the redshift space kernel including the standard Kaiser term~\cite{Kaiser:1987qv,Hamilton:1997zq}, Doppler~\cite{Raccanelli2016}, and relativistic contribution~\cite{Jeong2011} for the tracer~$X$ evaluated at~$\chi_i$, where $i=1,2$, 
\begin{equation}
    K_{X}(k,\mu,\chi_i) = \left[ b_{X,i} +f_i \mu^2 +\frac{\mathcal{A}_{X,i}}{k^2} \mp i\frac{f_i\mu\alpha_{X,i}}{k\chi_i} \right] \, ,
\end{equation}
and
\begin{equation}
    \mathcal{A}_{\mathrm{X},i}(b_{e,\mathrm{X}},Q_{\mathrm{X}},z_i) = \frac{H(z_i)^2}{(1+z_i)^2} \frac{3}{2}\Omega_m(z_i)\mathcal{B}(b_{e,\mathrm{X}},Q_{\mathrm{X}},z_i) \, .
\end{equation}
Note that, when building the two-point correlator, we assume the plane-parallel approximation.
An investigation on the range of validity of this approximation is discussed in Appendix~\ref{app:flatsky}, where we show that for the redshift range considered in Sect.~\ref{sec:relevance} we can fully trust this approach.
Introducing the mean distance~$\chi=(\chi_1+\chi_2)/2$ and the unequalness-in-time~$\delta\chi=\chi_1-\chi_2$, and changing variables from~$(\chi_1,\chi_2)$ to~$(\chi,\delta\chi)$, Equation~\eqref{eq:correlator} becomes
\begin{equation}
\begin{split}
    & \VEV{ \delta_{g,A}(\vec{\ell}_1,z_1[\bar{\chi_1}]) \delta_{g,B}^*(\vec{\ell}_2,z_2[\bar{\chi}_2]) } \\
    &= \int \frac{d\chi}{\chi^2} d\delta\chi W(\chi_1(\chi,\delta\chi),\bar{\chi}_1) W(\chi_2(\chi,\delta\chi),\bar{\chi}_2) (2\pi)^2 \delta_D^{(2)}\left(\vec{\ell}_1-\vec{\ell}_2-\frac{\delta\chi}{2\chi}\left(\vec{\ell}_1+\vec{\ell}_2\right)\right) \\
    & \qquad  D(z[\chi_1(\chi,\delta\chi)]) D(z[\chi_2(\chi,\delta\chi)]) \int \frac{dk_{\hat{n}}}{2\pi} e^{ik_{\hat{n}}\delta\chi} K_A(k,\mu,\chi_1(\chi,\delta\chi)) K_B(k,\mu,\chi_2(\chi,\delta\chi)) \mathcal{P}(k_{\hat{n}}\vhat{n},\vec{k}_{\perp},\chi,\delta\chi) \, .
\end{split}
\end{equation}
If~$\chi_1=\chi_2$, we recover the perfect equal-time case, where the angular power spectrum can be easily calculated using the well-known properties of the Bessel functions (see~\cite{Gao2023} for a detailed discussion, and~\cite{RVa} for an equal-time calculation). 
Moreover, the argument of the Dirac delta function picks up an extra term~$-\delta\chi(\vec{\ell}_1+\vec{\ell}_2)/2\chi$ 
with respect to the equal-time case, where it is only~$\vec{\ell}_1-\vec{\ell}_2$. 
Refs.~\cite{RVa,RVb} show that the off-diagonal part of the Dirac delta gives rise to negligible corrections to the equal-time angular power spectrum, with respect to the other unequal-time corrections that arise from the power spectrum and the overdensity kernels,\footnote{We show in Appendix~\ref{app:offdiagonal} that this conclusion still holds true when including Doppler terms and GR effects, which were not yet accounted for in Ref.~\cite{RVa}.} therefore we will drop them in the following, and we will use the standard~$\delta_D^{(2)}\left(\vec{\ell}_1-\vec{\ell}_2\right)$.\\

Assuming that the window functions are narrow enough, so that they can be approximated by Dirac deltas centered at the mean values~$\bar{\chi}_1$ and~$\bar{\chi}_2$, we have
\begin{equation}
\label{eq:def:utCell}
    \ClUT(\ell,\bar{\chi},\delta\chi) = \frac{1}{\bar{\chi}^2} \int \frac{dk_{\hat{n}}}{2\pi} e^{ik_{\hat{n}}\delta\chi} K_A(k_{\hat{n}}\hat{\vec{n}},\vec{\ell}/\bar{\chi},\bar{\chi},\delta\chi) K_B(k_{\hat{n}}\hat{\vec{n}},\vec{\ell}/\bar{\chi},\bar{\chi},\delta\chi) \mathcal{P}(k_{\hat{n}}\hat{\vec{n}},\vec{\ell}/\bar{\chi},\bar{\chi},\delta\chi) \, ,
\end{equation}
which is the flat-sky version of the full-sky unequal-time angular power spectrum.
We refer the reader to Ref.~\cite{Gao2023} for a detailed discussion of the mathematical connection between flat-sky and full-sky, in terms of asymptotic behavior of the spherical Bessel functions.

The equal-time, observed power spectrum is defined as~\cite{RVb}
\begin{equation}
\label{eq:def:Pobs}
    P_{\rm obs}\left(k_{\hat{n}}, \ell/\bar{\chi}, \bar{\chi}\right) = \bar{\chi}^2 \int d\delta\chi \, e^{-i\delta\chi k_{\hat{n}}} \ClUT(\ell,\bar{\chi},\delta\chi) \, .
\end{equation}
where~$k_{\hat{n}}$ represents the Fourier counterpart of the radial separation between the two tracers~$\delta\chi$.
In this fashion we can transfer the unequal-time information contained in the angular power spectrum~$\ClUT(\ell,\bar{\chi},\delta\chi)$ to the observed power spectrum in Fourier space.
In this transformation, not only the spatial separation but also any contribution arising from time evolution—through redshift dependence —is effectively mapped into comoving spatial coordinates. This follows from using light-cone coordinates, which ensure that time evolution and projection effects are incorporated into the spatial structure of the observed power spectrum. 

Since the distance between the observed system and the observer is much larger than the radial distance between the sources (i.e.,~$\delta\chi/\bar{\chi} \ll 1$), and since radial correlations die off very steeply as one departs from the equal-time case, we can now expand the matter power spectrum and the kernels of the galaxy overdensity field
around the equal-time case~$\delta\chi=0$, as
\begin{equation}
\label{eq:TaylorExpansion}
     K_A(k_{\hat{n}}\vhat{n},\vec{\ell}/\bar{\chi},\bar{\chi},\delta\chi) K_B(k_{\hat{n}}\vhat{n},\vec{\ell}/\bar{\chi},\bar{\chi},\delta\chi) \mathcal{P}( k_{\hat{n}}\vhat{n},\vec{\ell}/\bar{\chi},\bar{\chi},\delta\chi)  \\
     = D\left[z(\bar{\chi})\right]^2 \sum_{n=0}^{\infty} c_n(k,\mu,\bar{\chi}) H\left[z(\bar{\chi})\right]^n \left(\delta\chi\right)^n \mathcal{P}_0(k) \, ,
\end{equation}
where~$D\left[z(\bar{\chi})\right]$ is the growth factor at the mean comoving distance between the two sources (i.e.,~$(\chi_1+\chi_2)/2$), the coefficient~$c_0$ contains the equal-time kernels, and the~$c_n$ embed the unequal-time (UT) corrections, whose detailed expressions are computed in Section~\ref{sec:grterms}.
The quantity~$\mathcal{P}_0(k)$ is the theoretical dark matter equal-time linear power spectrum defined in Equation~\eqref{eq:phyper} at redshift~$z=0$. 
Then
\begin{equation}
\label{eq:Pobs_expansion}
\begin{split}
    P_{\rm obs}\left(k_{\hat{n}}, \ell/\bar{\chi}, \bar{\chi}\right) =& \int d\delta\chi \int \frac{dk_{\hat{n}}'}{2\pi} e^{i \delta\chi (k_{\hat{n}}'-k_{\hat{n}})} D\left[z(\bar{\chi})\right]^2 \sum_{n=0}^{\infty} c_n(k',\mu,\bar{\chi}) H\left[z(\bar{\chi})\right]^n \delta\chi^n \mathcal{P}_0(k') \\
    =& D\left[z(\bar{\chi})\right]^2 \sum_{n=0}^{\infty} \int d\delta\chi \int \frac{dk_{\hat{n}}'}{2\pi} \left[ \left( -i H\left[z(\bar{\chi})\right] \frac{d}{dk_{\hat{n}}'} \right)^n e^{i \delta\chi (k_{\hat{n}}'-k_{\hat{n}})} \right] c_n(k',\mu,\bar{\chi}) \mathcal{P}_0(k') \\
    =& D\left[z(\bar{\chi})\right]^2 \sum_{n=0}^{\infty} \left( i H\left[z(\bar{\chi})\right] \frac{d}{dk_{\hat{n}}} \right)^n \left[ c_n(k,\mu,\bar{\chi}) \mathcal{P}_0(k) \right] \, .
\end{split}
\end{equation}
Equation~\eqref{eq:Pobs_expansion} is our master equation, that will be used in the following to derive the unequal-time corrections to the power spectrum.


\subsection{General Relativistic and Doppler unequal-time coefficients}
\label{sec:grterms}

While the simplest $c_n$ coefficients were derived in~\cite{RVa, RVb}, here we add the contributions to~$c_n(k,\mu,\Bar{\chi})$ coming from Doppler terms and local general relativistic corrections; we start by expanding~$\alpha$ and~$\mathcal{A}$ around the equal-time case,~$\delta\chi=0$.
At first order,
\begin{equation}
    \frac{ \mathcal{A}\qty( z\qty[\chi\qty(\bar{\chi},\delta\chi)] ) }{ \mathcal{A}\qty( z\qty[\chi\qty(\bar{\chi},0)] ) } = 1 + H \bigg[2\bigg(\frac{H'}{H}-\frac{1}{1+z}\bigg)+\frac{\Omega_m'}{\Omega_m}+\frac{\mathcal{B}'}{\mathcal{B}}\bigg)\bigg] \dchi \delta\chi +\mathcal{O}(\delta \chi^2) \, ,
\end{equation}
where~$\mathcal{A}$ depends on the evolution bias~$b_e$ and magnification bias~$Q$, and 
\begin{equation}
    \frac{ \alpha\qty( z\qty[\chi\qty(\bar{\chi},\delta\chi)] ) }{ \alpha\qty( z\qty[\chi\qty(\bar{\chi},0)] ) } = 1 + H \bigg(\frac{H'}{H}-\frac{1}{1+z}+\frac{\Tilde{\alpha}'}{\Tilde{\alpha}}+\frac{1}{H\bar{\chi}}\bigg) \dchi \delta\chi +\mathcal{O}(\delta \chi^2) \, ,
\end{equation}
with~$\alpha$ being the Doppler term coefficient, and where a prime denotes derivative with respect to redshift.
We have rewritten the Doppler term in Equation~\eqref{eq:alpha} as~$\alpha=-H\chi \Tilde{\alpha} / (1+z)$, with
\begin{equation}
    \Tilde{\alpha}=b_e(z) -1 -2 Q(z) +\frac{3}{2} \Omega_m(z) -\frac{2(1-Q(z))(1+z)}{H(z)\chi(z)}\,.
\end{equation}
Explicitly,
\begin{equation}
    \Tilde{\alpha}'=b_e'-2 Q'+\frac{3}{2} \Omega_m'+2\qty[\frac{Q' (1+z)}{H\bar{\chi}}+\frac{(1-Q)(1+z)}{(H\bar{\chi})^2}-\frac{1-Q}{H\bar{\chi}} +\frac{(1-Q)(1+z)}{H\bar{\chi}} \frac{H'}{H}] \, ,
\end{equation}
and
\begin{equation}
\begin{split}
    \mathcal{B}' =& b_e' \left( 1-\frac{2f}{3\Omega_m} \right) -\frac{2 b_e f}{3 \Omega_m} \left( \frac{f'}{f} -\frac{\Omega_m'}{\Omega_m} \right) +\frac{2f}{\Omega_m} \left( \frac{f'}{f} -\frac{\Omega_m'}{\Omega_m} \right) +\frac{3}{2}\Omega_m' -4Q' -f' \\
    & +\frac{2}{H\bar{\chi}} \left[ Q' (1+z) -(1-Q) \right] +\frac{2(1-Q)(1+z)}{H\bar{\chi}} \left( \frac{H'}{H} -\frac{1}{H\bar{\chi}} \right) \, .
\end{split}
\end{equation}
Isolating the dependence on the wave-vector and orientation angle, the first-order coefficient in Equation~\eqref{eq:Pobs_expansion} is
\begin{equation}
\begin{split}
    c_1(k,\mu,\bar{z}) =& c_{100}(\bar{z}) +\mu^2 c_{120}(\bar{z}) +\mu^4 c_{140}(\bar{z}) +\left(\frac{H}{k}\right)^2 c_{102}(\bar{z}) +\mu^2 \left(\frac{H}{k}\right)^2 c_{122}(\bar{z}) +\left(\frac{H}{k}\right)^4 c_{104}(\bar{z}) \\
    & +i \mu\left(\frac{H}{k}\right) c_{111}(\bar{z}) +i \mu^3 \left(\frac{H}{k}\right) c_{131}(\bar{z}) +i \mu \left(\frac{H}{k}\right)^3 c_{113}(\bar{z}) \, .
\end{split}
\end{equation}

At this stage, different choices of the mean distance are possible -- see Appendix B of~\cite{RVa} for more details.
As in the previous section, we choose the arithmetic mean,~$\Bar{\chi}=(\chi_A+\chi_B)/2$, so that~$\delta\chi=\chi_A-\chi_B$ and~$d\chi_{A,B} / d\delta\chi = \pm 1/2$. 
The coefficients are then
\begin{subequations}
\label{eq:c1ij_real}
\begin{align}
    c_{100} =&\frac{1}{2}  b_A b_B \left( \frac{b_A'}{b_A}-\frac{b_B'}{b_B} \right) \, , \\
    c_{120} =&  \frac{f}{2}\left[ b_A' -b_B' -\frac{f'}{f}\left(b_A-b_B\right) \right] \, , \\
    c_{140} =& 0 \, , \\
    c_{102} =&  \frac{1}{2}\left[ \left( b_A' \frac{\mathcal{A}_B}{H^2} -b_A \frac{\mathcal{A}_B'}{H^2} \right) -\left( b_B' \frac{\mathcal{A}_A}{H^2} -b_B \frac{\mathcal{A}_A'}{H^2} \right) \right] \, , \\
    c_{122} =& \frac{f}{2}\left[ \frac{\mathcal{A}_A'}{H^2}-\frac{\mathcal{A}_B'}{H^2} -\frac{f'}{f}\left(\frac{\mathcal{A}_A}{H^2}-\frac{\mathcal{A}_B}{H^2}\right) +\frac{f \alpha_A \alpha_B}{(H\chi)^2} \left( \frac{\alpha_A'}{\alpha_A}-\frac{\alpha_B'}{\alpha_B} \right) \right] \, , \\
    c_{104} =&\frac{1}{2} \frac{\mathcal{A}_A}{H^2} \frac{\mathcal{A}_B}{H^2} \left( \frac{\mathcal{A}_A'}{\mathcal{A}_A} -\frac{\mathcal{A}_B'}{\mathcal{A}_B} \right) \, ;
\end{align}
\end{subequations}
\begin{subequations}
\label{eq:c1ij_imaginary}
\begin{align}
    c_{111} =& \frac{f}{2 H\chi} \left[ b_A'\alpha_B+b_B'\alpha_A -b_A\alpha_B' -b_B\alpha_A' +\left(b_A\alpha_B+b_B\alpha_A\right) \left(\frac{1}{H\chi}-\frac{f'}{f}\right) \right] \, , \\
    c_{131} =& \frac{f^2}{ 2H\chi} \left( \frac{\alpha_A+\alpha_B}{H\chi} -\alpha_A' -\alpha_B' \right) \, , \\
    c_{113} =& \frac{f}{ 2H\chi} \left[ \left(\alpha_A\frac{\mathcal{A}_B}{H^2}+\alpha_B\frac{\mathcal{A}_A}{H^2} \right)\left(\frac{1}{H\chi}-\frac{f'}{f}\right) +\alpha_A\frac{\mathcal{A}_B'}{H^2} +\alpha_B\frac{\mathcal{A}_A'}{H^2} -\alpha_A'\frac{\mathcal{A}_B}{H^2} -\alpha_B'\frac{\mathcal{A}_A}{H^2} \right] \, .
\end{align}
\end{subequations}
The first set of coefficients, the~$c_{1ij}$ with~$\{i,j\}$ being even, gives rise to an imaginary unequal-time correction. The second set, the~$c_{1ij}$ with~$\{i,j\}$ being odd, gives rise to a real correction.
All redshift-dependent quantities in Equations~\eqref{eq:c1ij_real}--~\eqref{eq:c1ij_imaginary} are evaluated at~$\delta\chi=0$, by construction.
With the arithmetic choice of mean distance, the coefficient~$c_{140}$ vanishes since it only receives contributions from the~$f\mu^2$ terms of the kernels.

The terms proportional to~$\mathcal{A}'$ and~$\alpha'$ depend on the time derivatives of the evolution and magnification biases,~$b_e'$ and~$Q'$, that are related to the galaxies' redshift and luminosity distribution.

In the absence of Doppler and general relativistic terms, as noted in~\cite{RVa,RVb}, the first order unequal-time correction would vanish in the single-tracer case, while in the multi-tracer case it would be purely imaginary, and sourced by the first set of coefficients.
When including Doppler terms and general relativistic corrections, instead, the second set of coefficients is non-vanishing, even in the single-tracer case, and it gives rise to a real first order correction as well.
Explicitly, in the single-tracer case, the first set of coefficients, Equations~\eqref{eq:c1ij_real}, vanishes, and Equations~\eqref{eq:c1ij_imaginary} become
\begin{subequations}
\begin{align}
c_{111} =& \frac{f b \alpha}{H\chi} \left[ \frac{b'}{b} -\frac{\alpha'}{\alpha} +\left(\frac{1}{H\chi}-\frac{f'}{f}\right) \right] \, , \\
c_{131} =& \frac{f^2 \alpha}{H\chi} \left( \frac{1}{H\chi} -\frac{\alpha'}{\alpha} \right) \, , \\
c_{113} =& \frac{f \alpha}{H\chi} \frac{\mathcal{A}}{H^2} \left[ \left(\frac{1}{H\chi}-\frac{f'}{f}\right) -\frac{\alpha'}{\alpha} +\frac{\mathcal{A}'}{\mathcal{A}} \right] \, .
\end{align}
\end{subequations}

\begin{figure}
\centering   
\includegraphics[width=1\linewidth]{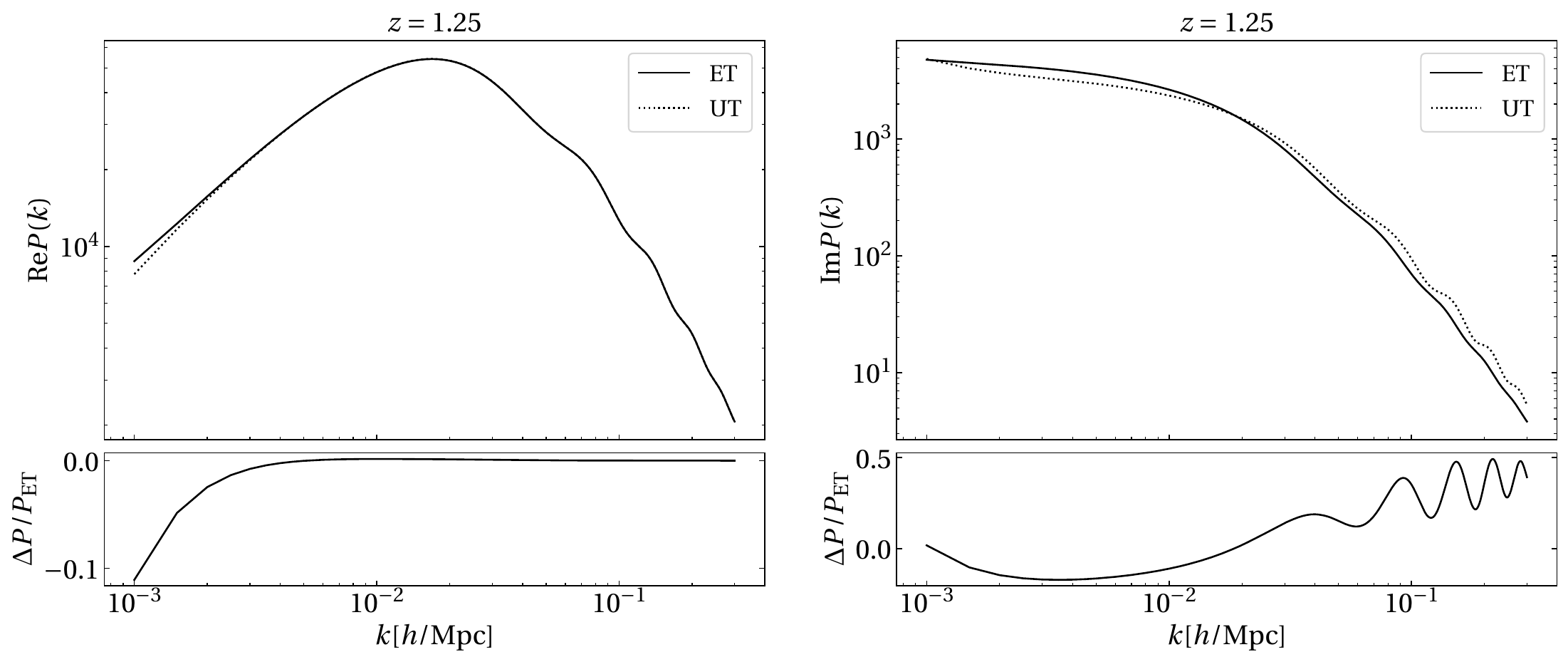}
\caption{Real (\textit{left}) and imaginary (\textit{right}) part of the galaxy power spectrum in the equal-time and unequal-time case for~$\mu=1$. The residuals for UT corrections~$\Delta P$ normalized to the ET power spectrum are shown in the lower panels.}
\label{fig:etut}
\end{figure}

In general, the largest deviations coming from UT contributions are imaginary, as illustrated in Figure~\ref{fig:etut}.
There we show the real and imaginary part of the redshift-space galaxy power spectrum in the equal-time and unequal-time cases\footnote{The imaginary part of the equal-time power spectrum is zero in the single-tracer case, but when cross-correlating two different tracers, the equal-time power spectrum picks up a non-vanishing imaginary part (only when Doppler terms are included).}, for~$\mu=1$ (as unequal-time corrections are expected to be maximized for radial correlators).
As an example case, we used~$b_A=(2+z)^{0.5}$, $b_B=(0.5+z)^{1.2}$, and number counts distributions~$\frac{dN_X}{dz}\propto z^{\gamma_X}e^{-(z/0.5)^{1.5}}$ with~$\gamma_A=4,\gamma_B=1.1$.
The relative importance of UT corrections (i.e.,~$(P_{\mathrm{UT}}-P_{\mathrm{ET}})/P_{\mathrm{ET}}$), is shown in the lower panels, where one can see that the bulk of the UT contribution is in the imaginary part of the multi-tracer power spectrum.
Such imaginary corrections arise because in general the multi-tracer power spectrum is not invariant under the transformation~$\mu\rightarrow-\mu$.

\subsection{Primordial non-Gaussianity}
\label{sec:PnG}
Since UT corrections can give a contribution to the power spectrum at large scales, it is natural to wonder whether there might be degeneracies with the imprints of local-type Primordial non-Gaussianity (PnG).
A similar investigation was carried out in the context of Doppler and relativistic equal-time corrections, in~\cite{Maartens2012, Raccanelli2016}).
In order to do this, the first step is to derive the unequal-time contributions in the presence of Primordial non-Gaussianity.

The deviation from Gaussian primordial curvature perturbations can be parameterized by the dimensionless parameter~$\fnl$  as~\cite{Salopek1990, Komatsu2001}
\begin{equation}
\label{eq:fnl}
    \Phi=\phi+\fnl\left[\phi^2-\langle\phi^2\rangle\right] \, ,
\end{equation}
where~$\Phi$ is the Bardeen's gauge-invariant potential, which on sub-horizon scales coincides with the Newtonian gravitational potential, and~$\phi$ is its Gaussian part.
Local-type PnG induces a long-short mode coupling that affects halo clustering at late time, through a scale-dependent modification of the halo bias~\cite{Matarrese2008, Dalal2007, Slosar2008, Verde2009, Desjacques2010}
\begin{equation}
\label{eq:ng-bias}
    b(k,z) = b_{\rm G}(z) + b_{\rm NG}(k,z)=b_{\rm G}(z)+ \fnl b_{\phi}(z) \frac{3\Omega_{m,0} H_0^2}{2 k^2T(k) D(z)} \, ,
\end{equation}
where~$b_{\rm G}(z)$ is the Gaussian linear bias,~$T(k)$ is the matter transfer function.
Expanding the non-Gaussian bias around~$\delta\chi=0$ leads to
\begin{equation}
     \frac{ b_{\rm NG}\qty( z\qty[\chi\qty(\bar{\chi},\delta\chi)] ) }{ b_{\rm NG}\qty( z\qty[\chi\qty(\bar{\chi},0)] ) } = 1 + H\bigg( \frac{b_{\phi}'}{b_{\phi}}+\frac{f}{1+z} \bigg) \dchi \delta\chi +\mathcal{O}(\delta \chi^2) \, ,
\end{equation}
which gives the additional UT correction
\begin{equation}
\begin{gathered}
c_{1,\rm NG}(k,\mu,\bar{z})\hspace{-0.11cm}=\hspace{-0.1cm}\frac{3\fnl\Omega_{m,0} H_0^2}{2 T(k) D_{\rm md}(z)} \left[\hspace{-0.13cm}\left(\frac{H}{k}\right)^2 \hspace{-0.2cm} c_{102,\rm NG}(\bar{z})+\mu^2\left(\frac{H}{k}\right)^2 \hspace{-0.2cm} c_{122,\rm NG}(\bar{z})+\left(\frac{H}{k}\right)^4 \hspace{-0.2cm} c_{104,\rm NG}(\bar{z})+i\mu\left(\frac{H}{k}\right)^3 \hspace{-0.2cm} c_{113,\rm NG}(\bar{z})\right] ,
    \end{gathered}
\end{equation}
where the coefficients are
\begin{subequations}
\begin{align}
c_{102,\rm NG}=&\frac{1}{2H^2}\left[\frac{f}{1+z}(b_{\phi,A}b_B-b_{\phi,B}b_A)+b_{\phi,B}b_A'-b_{\phi,A}b_B'+b_{\phi,A}'b_B-b_{\phi,B}'b_A\right] \, , \\
c_{122,\rm NG}=&\frac{f}{2H^2}\left[\frac{f'}{f}(b_{\phi,B}-b_{\phi,A})+b_{\phi,A}'-b_{\phi,B}'+\frac{f}{1+z}(b_{\phi,A}-b_{\phi,B})\right] \, , \\
c_{104,\rm NG}=&\frac{1}{2H^4}\left[b_{\phi,B}\mathcal{A}_A'- b_{\phi,A}\mathcal{A}_B'+\frac{f}{1+z}(\mathcal{A}_Bb_{\phi,A}- \mathcal{A}_Ab_{\phi,B})+\mathcal{A}_Bb_{\phi,A}'-\mathcal{A}_Ab_{\phi,B}'\right] \, , \\
c_{113,\rm NG}=&\frac{f}{2H^3\chi}\bigg[(\alpha_Bb_{\phi,A}'+\alpha_Ab_{\phi,B}')+\frac{f}{1+z}(\alpha_Ab_{\phi,B}+\alpha_Bb_{\phi,A}) \\
& \qquad +\frac{1}{H\chi}(\alpha_Ab_{\phi,B}+\alpha_Bb_{\phi,A})+\frac{f'}{f}(-\alpha_Ab_{\phi,B}-\alpha_Bb_{\phi,A})-\alpha_A'b_{\phi,B}-\alpha_B'b_{\phi,A}\bigg] \, .
\end{align}
\end{subequations}

For simplicity, we assume the universality relation, so that~$b_{\phi}=2\delta_c(b_{\rm G}-1)$, where~$\delta_c=1.686$ is the critical value of the matter overdensity for spherical collapse (for some generalizations see, e.g.,~\cite{Sheth:1999su,Pillepich:2008ka,Grossi:2009an,Fondi:2023egm}).

Even if the leading unequal time corrections are imaginary, one can see from Figure~\ref{fig:etut} that the real part of the UT contributions can give an $\sim10\%$ correction to the standard equal time power spectrum at large scales. It is interesting to investigate whether this is degenerate with contributions coming from the scale-dependent bias sourced by a non-zero $\fnl$. To proceed, let us write the main real UT contributions to the ET power spectrum.
Recall that in general
\begin{equation}
    \Delta_{\mathrm{UT}}\propto\frac{d}{dk_{\hat{n}}}\left[c_{1}(k,\mu,\bar{\chi}) \mathcal{P}_0(k) \right]\propto\frac{d}{dk_{\hat{n}}}\left[\sum_{a,b} H^b\frac{\mu^a}{k^b}c_{1ab}(\bar{\chi}) \mathcal{P}_0(k)\right] \, ,
\end{equation}
where~$c_{1ab}$ are given by Equations~\eqref{eq:c1ij_real} and Equations~\eqref{eq:c1ij_imaginary}.
Since~$k$ and~$\mu$ both depend on~$k_{\hat{n}}$,
\begin{equation}
    \frac{d}{dk_{\hat{n}}} \left[\frac{\mu^a}{k^b}c_{1ab}(\bar{\chi}) \mathcal{P}_0(k)\right]=\bigg\{\frac{1}{k^{b+1}}\big[-b\mu^{a+1}+a\mu^{a-1}(1-\mu^2)\big]c_{1ab}(\bar{\chi}) \mathcal{P}_0(k)+  c_{1ab}(\bar{\chi})\frac{\mu^{a+1}}{k^b}\frac{d\mathcal{P}_0(k)}{dk}\bigg\} \, .
\end{equation}
 
\begin{figure}
\centering
\includegraphics[width=1\linewidth]{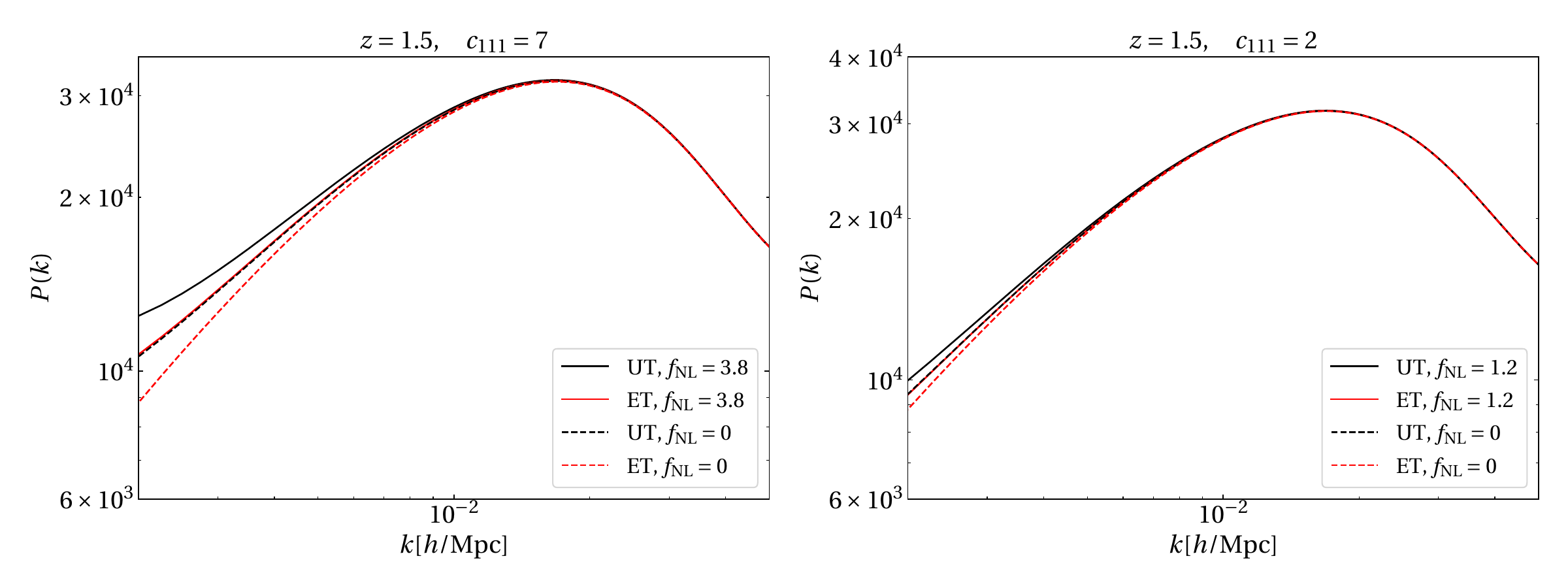}
\includegraphics[width=1\linewidth]{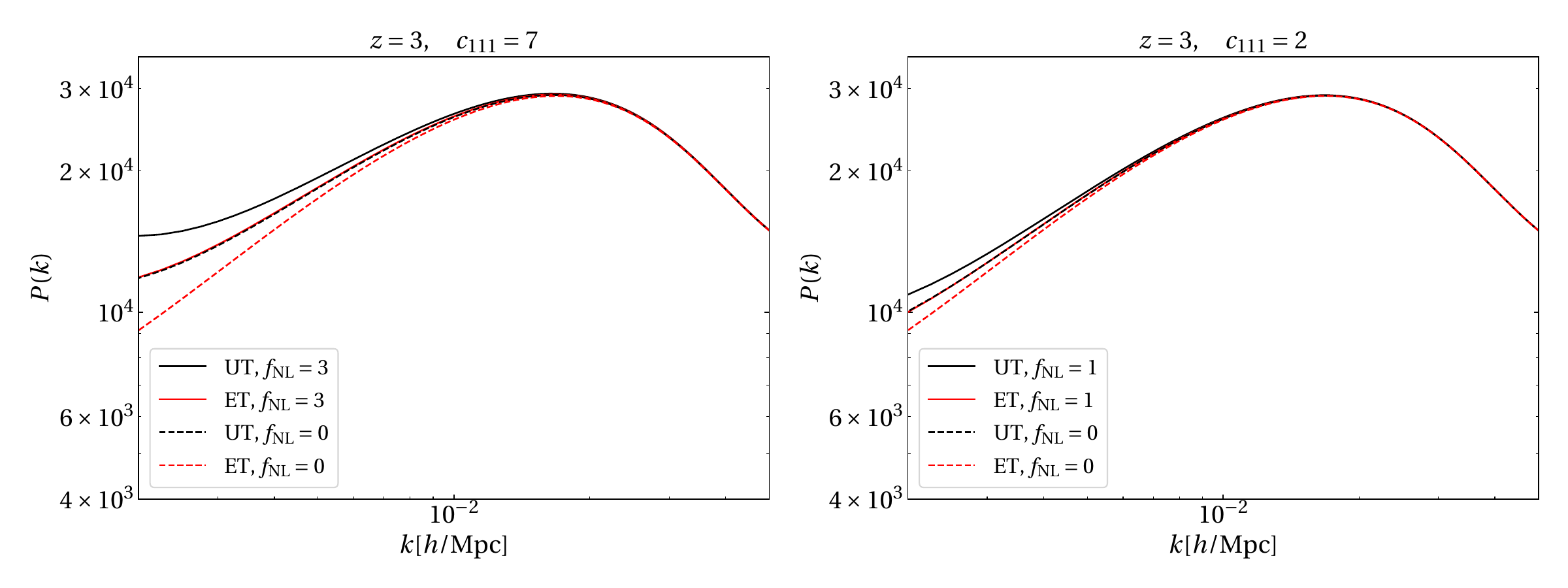}
\caption{Equal-time power spectrum and unequal-time power spectrum for different values of the coefficient~$c_{111}$ at redshift~$z=1.5$ (\textit{upper panels}) and~$z=3$ (\textit{lower panels)} at~$\mu=0.4$. Here~$c_{131}=1$.}
\label{fig:fnl}
\end{figure}

On very large scales~$d\mathcal{P}_0(k)/dk\sim \mathcal{P}_0(k)/k$; as shown explicitly in Appendix~\ref{app:coeffs}, the dominant real UT contributions are sourced by the coefficients~$c_{111},c_{131}$, therefore
\begin{equation}
    \Delta_{UT,\mathrm{real}} \sim H^2\left(1-\mu^2\right)(c_{111}(\bar{\chi})+3\mu^2c_{131}(\bar{\chi}))\frac{\mathcal{P}_0(k)}{k^2} \, .
\end{equation}
One can notice that the real part of the UT corrections exhibits the same~$k^{-2}$ behavior of the scale dependent non-Gaussian bias of Equation~\eqref{eq:ng-bias} indicating that these kinds of corrections in principle can mimic an effective~$\fnl$ on large scales. 
In general the precise value of the effective~$\fnl$  depends on the values of the dimensionless coefficients~$c_{111},c_{131}$, which are quantities of order one and depend on~$\{b_A,b_B,\alpha_A,\alpha_B,f\}$ and their derivatives. As a first investigation, in Figure~\ref{fig:fnl} we present some values of~$c_{111},c_{131}$  which can mimic different values of~$\fnl$ at different redshift, for~$\mu=0.4$ as an example (in general the net effect exhibits a weak dependence on~$\mu$). Non-Gaussian power spectra are shown for~$\fnl=\hat{f}_{\mathrm{NL}}$ where~$\hat{f}_{\mathrm{NL}}$ is the effective~$\fnl$ mimicked by UT corrections. Notably, ET power spectra with~$\fnl=\hat{f}_{\mathrm{NL}}$ (red dashed) overlap with the UT models with zero~$\fnl$ (dashed black) showing that~$\hat{f}_{\mathrm{NL}}$ is indeed the effective~$\fnl$  mimicked by unequal-time corrections. This effects varies non-trivially with~$z$ depending on the specific values of the biases and~$\alpha$ of the tracers considered.
While we leave a detailed investigation of the impact of this for future surveys, we can notice that the effective~$\fnl$ mimicked by the UT corrections seems to be relevant for the targets of current and future galaxy surveys measuring PnG.

\section{Relevance of unequal-time corrections}
\label{sec:relevance}
We present here a first investigation of the relevance of first-order unequal-time corrections to the galaxy power spectrum.
Throughout the analysis, we use \textit{Planck} 2018 fiducial values for the cosmological parameters~\cite{Planck2018} and a fiducial of~$\fnl=1$, and checked that a different choice of the fiducial value of~$\fnl$ in the range allowed by the latest CMB constraints~\cite{Planck:2019kim} would not greatly impact the following results.

In order to estimate quantitatively the relevance of all UT corrections, we compute their total signal-to-noise ratio (SNR). 
In the multi-tracer case, the power spectrum covariance is~\cite{McDonald:2008sh,Seljak:2008xr}
\begin{equation}
\label{eq:covMT}
    \text{Cov}(\vec{k},\bar{z}) = \frac{4\pi^2}{V_s(\bar{z}) k^2\Delta k\Delta\mu} \left[ P_{{\rm obs},AA}^{\rm tot}(\vec{k},\bar{z}) P_{{\rm obs},BB}^{\rm tot}(\vec{k},\bar{z}) +\left( P_{{\rm obs},AB}^{\rm tot}(\vec{k},\bar{z}) \right)^2 \right] \, ,
\end{equation}
where~$P_{{\rm obs},XY}^{\rm tot} = P_{{\rm obs},XY} +\delta^K_{XY} / \Bar{n}_X$, and we neglected cross-shot noise terms since we assume that the stochastic components of two galaxy samples are statistically independent.
At first order in~$\delta\chi$,
\begin{equation}
\label{eq:gencov}
\begin{aligned}
    \mathrm{Cov}(\vec{k},\bar{z}) \simeq &\frac{4\pi^2}{V_s(\bar{z}) k^2\Delta k\Delta\mu}\bigg\{ P_{\mathrm{ET},AB}(\vec{k},\bar{z})^2 + \left(P_{\mathrm{ET},AA}(\vec{k},\bar{z}) +\frac{1}{\bar{n}_A}\right) \left(P_{\mathrm{ET},BB}(\vec{k},\bar{z}) +\frac{1}{\bar{n}_B}\right) \\
    &+2 P_{\mathrm{ET},AB}(\vec{k},\bar{z}) D(\bar{z})^2 i H(\bar{z}) \partial_{k_{\hat{n}}}\left[c_{1,AB}(k,\mu,\bar{z})\mathcal{P}_0(k)\right] \bigg\}\, .
\end{aligned}
\end{equation}
where~$P_{\mathrm{ET},XY}(\vec{k},\bar{z})=D(\bar{z})^2c_{\mathrm{0},XY}(k,\mu,\bar{z})\mathcal{P}_0(k)$, and~$V_s(\bar{z})$ is the volume of the redshift bin. 
Note that we neglected UT contributions coming from the single-tracer galaxy power spectrum. The resulting signal-to-noise ratio is therefore
\begin{equation}
\label{eq:SNR}
    {\rm SNR}^2(\bar{z}) = \int_{k,\mu} \frac{\left| D(\bar{z})^2 iH(\bar{z}) \partial_{k_{\hat{n}}}\left[c_{1,AB}(k,\mu,\bar{z})\mathcal{P}_0(k)\right] \right|^2}{\mathrm{Cov}(\vec{k},\bar{z})} \, .
\end{equation}
The UT covariance contains imaginary parts, but their contribution averages to zero when we perform the integration over the orientation angle, since the integrands are proportional to odd powers of~$\mu$.
It can be shown that, for the redshift range considered in this work, employing the equal-time version of the covariance gives an almost identical SNR.

Furthermore, unequal-time corrections become more important at high redshift.
As previously stated (and explicitly shown in Appendix~\ref{app:coeffs}), the dominant UT corrections are imaginary, and come from the coefficients~$c_{100}$.
So, to get a first idea on the behavior of the dominant contribution, UT corrections can be approximated as
\begin{equation}
    \Delta_{\rm UT}\sim i \mu H(\bar{z})c_{100}(\bar{z})\frac{d\mathcal{P}_0(k)}{dk} \, ,
\end{equation}
where recall that~$c_{100}=\frac{1}{2}  b_A b_B \left( \frac{b_A'}{b_A}-\frac{b_B'}{b_B} \right)$. 
Note that the scale dependence of the imaginary corrections is governed by the derivative of the linear power spectrum~$d\mathcal{P}_0/dk$; this leads to a sign change around the matter-radiation equality scale, as one can see in the left panel of Figure~\ref{fig:etut}.
As an illustrative toy model, one can approximate the biases of the two tracers as some power law, i.e.,~$b_A^*(1+z)^{\alpha}, b_B^*(1+z)^{\beta}$; then 
\begin{equation}
    \Delta_{\rm UT}\sim \frac{i}{2} \mu aH(\bar{z}) \frac{d\mathcal{P}_0(k)}{dk} b_A^*b_B^* (\alpha-\beta) \, ,
\end{equation}
which increases for correlations along the line of sight -- as expected, since UT effects are enhanced when the radial separation between the two sources is maximized -- and at high redshifts.

\begin{figure}
\centering
\includegraphics[width=0.49\linewidth]{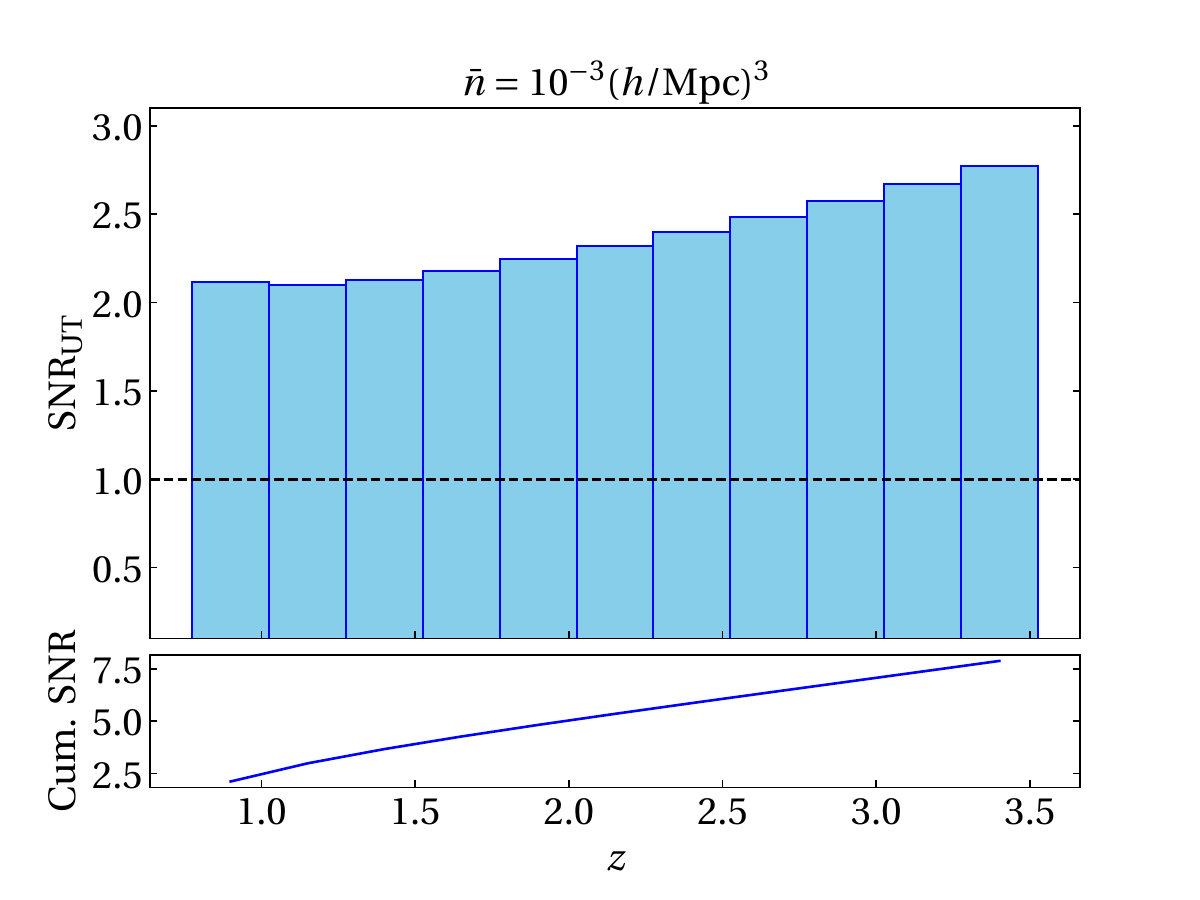}
\includegraphics[width=0.49\linewidth]{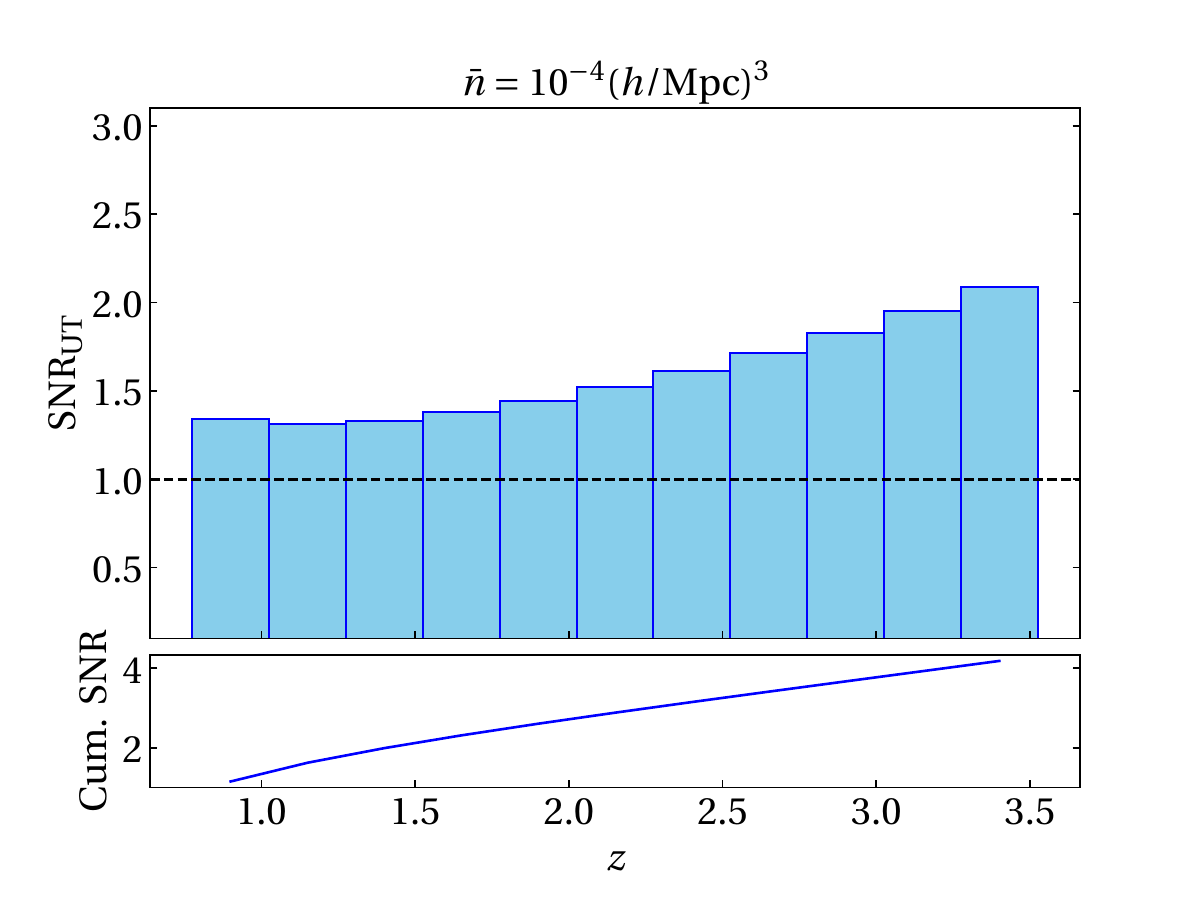}
\caption{SNR of UT contributions for bins centered at mean redshifts~$z$, for~$\bar{n}=10^{-3} \, (h/\mathrm{Mpc})^3$ (\textit{left}) and~$\bar{n}=10^{-4} \, (h/\mathrm{Mpc})^3$ (\textit{right}), using~$b_A=(2+z)^{0.5}$ and~$b_B=(0.5+z)^{1.2}$. The bottom panel shows the cumulative SNR when including higher~$z$ bins.}
\label{fig:snr_plot}
\end{figure}

This is confirmed in Figure~\ref{fig:snr_plot}, where the signal-to-noise ratio of unequal-time corrections is shown for different redshift bins centered at redshift~$z$, for two different fixed galaxy number densities~$\bar{n}\sim10^{-3} \, ({\rm Mpc}/h)^{-3}$ (left panel) and ~$\bar{n}\sim10^{-4} \, ({\rm Mpc}/h)^{-4}$ (right panel).
We assume~$f_{\rm sky}\sim0.7$, biases~$b_A=(2+z)^{0.5}$ and~$b_B=(0.5+z)^{1.2}$, and bins with the same comoving volume of~$V=(3800 \, \mathrm{Mpc}/h)^3$.

\begin{figure}
\centering
\includegraphics[width=1\linewidth]{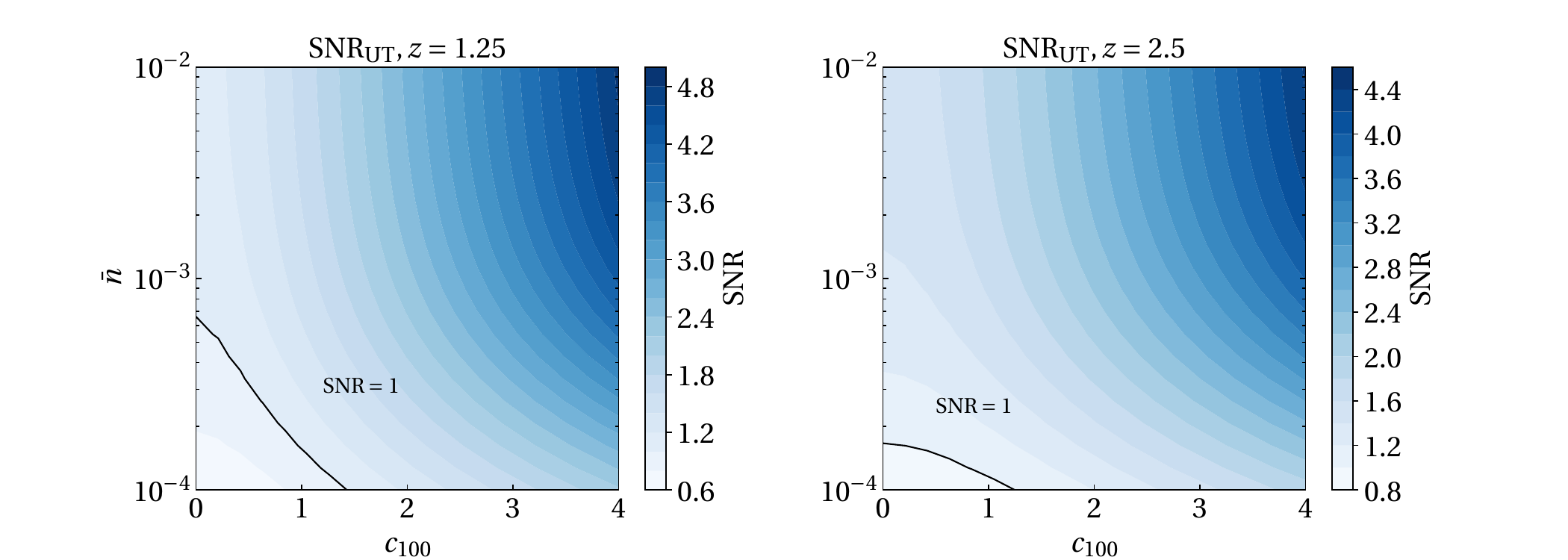}
\caption{SNR of UT corrections, at~$\bar{z}=1.25$ (\textit{left)} and~$\bar{z}=2.5$ (\textit{right)}, as a function of the mean number density of galaxies~$\bar{n}$ and of the~$c_{100}$ coefficient, which depends on the difference between the biases and on their derivatives.}
    \label{fig:snr_plot1}
\end{figure}

As an example, we investigate a Stage IV-{\it like} scenario, combining two redshift bins in the range~$1<z<2$; in this case, one obtains a signal-to-noise-ratio for the UT corrections $\text{SNR}\sim3$.
As shown explicitly~in Appendix~\ref{app:flatsky}, our flat-sky model can be fully trusted at~$z\gtrsim1$ where the curvature of the sky can be neglected in the theoretical modeling.
Furthermore, UT corrections are sensitive to the biases of the two tracers and their redshift evolution through the coefficient~$c_{100}$, which is in general a quantity of order one.
This is illustrated in Figure~\ref{fig:snr_plot1}, where we show the dependence of the SNR on~$c_{100}$ and on the number density of galaxies, for redshift bins centered at~$z=1.25$ and~$z=2.5$. Finally, note that Doppler terms source a first-order real unequal-time contribution to the standard equal-time power spectrum, which is not vanishing also in the single-tracer case.
The expression of this correction can be found in Appendix~\ref{app:single}. However, it can be shown that UT corrections in the single-tracer case are negligible for any realistic Stage IV-\textit{like} scenario.

\section{Conclusions}
\label{sec:conclusions}

In this work, we present a generalization of the formalism for the unequal-time (UT) observed power spectrum, which includes the contributions from local-type primordial non-Gaussianity, Doppler terms, and local general relativistic corrections~\footnote{Note that we do not include the contribution of the observer’s velocity to the imaginary part of the power spectrum, as discussed in~\cite{Bertacca2019, Bahr-Kalus:2021jvu, Elkhashab:2021lsk, Elkhashab:2024yup}}.
The unequal-time framework extends the standard power spectrum analysis by introducing corrections that accounts for radial information, which is otherwise lost in standard equal-time treatments~\cite{RVa, RVb, Steele:2025djd}.

In the Newtonian limit and when Doppler terms are neglected, the redshift-space galaxy power spectrum receives a first-order UT correction which is purely imaginary and non-vanishing only in the multi-tracer case.
This correction arises because the multi-tracer power spectrum is not invariant under the transformation~$\mu\rightarrow-\mu$.
Such contributions represent the dominant unequal-time correction to the standard equal-time power spectrum, and they are expected to be more important when correlating tracers with very different biases.
In fact, they strongly depend on the redshift evolution of the linear biases of the two tracers. Moreover, these corrections are more important at high redshift and are maximized for radial correlations.

When including Doppler terms, the first-order UT corrections acquire a real part (that is non-vanishing even in the single-tracer scenario), which becomes non negligible at very large scales and exhibits the same $k^{-2}$ behavior as the scale-dependent bias sourced by a non-zero $\fnl$. Indeed, we find that the real part of UT corrections can potentially mimic an effective $\fnl$ of order one -- the precise value depending on the two tracers and on the redshift considered -- which could be relevant for the targets of current and future galaxy surveys measuring primordial non-Gaussianity. A more rigorous analysis on the bias induced by UT corrections when measuring $\fnl$ is the subject of an ongoing work.

Furthermore,we investigate the relevance of UT corrections, showing their behavior for different redshifts and biasing models. We find that such corrections can become relevant and detectable (or large enough to need to be corrected), depending, other than the bias models, on the redshift range and source number density of the survey.

To have a first estimate for a realistic situation, we look at a Stage IV-{\it like} galaxy survey, for which we obtain a SNR of the UT corrections of order~$\sim3$.

While the exact value depend on the multi-tracer situation, this shows that such contributions are potentially at least non negligible, and will need to be accounted in future survey analyses.

Our results are a first demonstration of the relevance of UT corrections for future galaxy surveys, and they motivate further investigation of their impact on the modeling of galaxy clustering.
A more rigorous evaluation of the impact of UT corrections on constraining cosmological parameters, structure growth, and local primordial non-Gaussianity, as well as a possible extension of this formalism at lower redshifts, are the subjects of other papers in preparation.

\begin{acknowledgments}
The authors thank Nicola Bellomo, Federico Semenzato, Theo Steele for valuable discussions. AR acknowledges funding from the Italian Ministry of University and Research
(MUR) through the ``Dipartimenti di eccellenza'' project ``Science of the Universe''. DB acknowledges support from the COSMOS network (www.cosmosnet.it) through the ASI (Italian Space Agency). Z.V. acknowledges the support of the Kavli Foundation.
\end{acknowledgments}

\bibliography{bibliography}

\appendix
\section{Impact of the flat sky approximation}
\label{app:flatsky}

Recall that in Equation~\eqref{eq:def:Pobs} we have defined the observed power spectrum as
\begin{equation}
\begin{split}
    P_{\rm obs}\left(k_{\hat{n}}, \ell/\bar{\chi}, \bar{\chi}\right) &= \bar{\chi}^2 \int d\delta\chi \, e^{-i\delta\chi k_{\hat{n}}} \ClUT(\ell,\bar{\chi},\delta\chi) \\
    &=D\left[z(\bar{\chi})\right]^2 \sum_{n=0}^{\infty}  \left( i H\left[z(\bar{\chi})\right] \frac{d}{dk_{\hat{n}}} \right)^n[c_{n}(k_{\hat{n}},\ell/\bar{\chi},z(\bar{\chi}))\mathcal{P}_0(k_{\hat{n}},\ell/\bar{\chi},z(\bar{\chi})] \, .
\end{split}
\end{equation}
where~$\ClUT(\ell,\bar{\chi},\delta\chi)$ is the unequal-time flat-sky angular power spectrum.
It is interesting to see what happens if we substitute in the integral the full-sky angular power spectrum~$C_{\ell}(\Bar{\chi},\delta\chi)$. This comparison can give us a hint of the validity of the flat sky approximation. 
Then the two models to be compared are
\begin{equation}
\begin{gathered}
    P_{\rm full sky}\left(k_{\hat{n}}, \ell/\bar{\chi}, \bar{\chi}\right) = \bar{\chi}^2 \int d\delta\chi \, e^{-i\delta\chi k_{\hat{n}}} C_{\ell}(\bar{\chi},\delta\chi) \, , \\
    P_{\rm flat sky}\left(k_{\hat{n}}, \ell/\bar{\chi}, \bar{\chi}\right) =D\left[z(\bar{\chi})\right]^2 \sum_{n=0}^{\infty}  \left( i H\left[z(\bar{\chi})\right] \frac{d}{dk_{\hat{n}}} \right)^n \bigg[c_{n}(k_{\hat{n}},\ell/\bar{\chi},z(\bar{\chi}))\mathcal{P}_0(k_{\hat{n}},\ell/\bar{\chi},z(\bar{\chi}))\bigg] \, .
\end{gathered}
\end{equation}

\begin{figure}
    \centering         
    \includegraphics[width=0.32\linewidth]{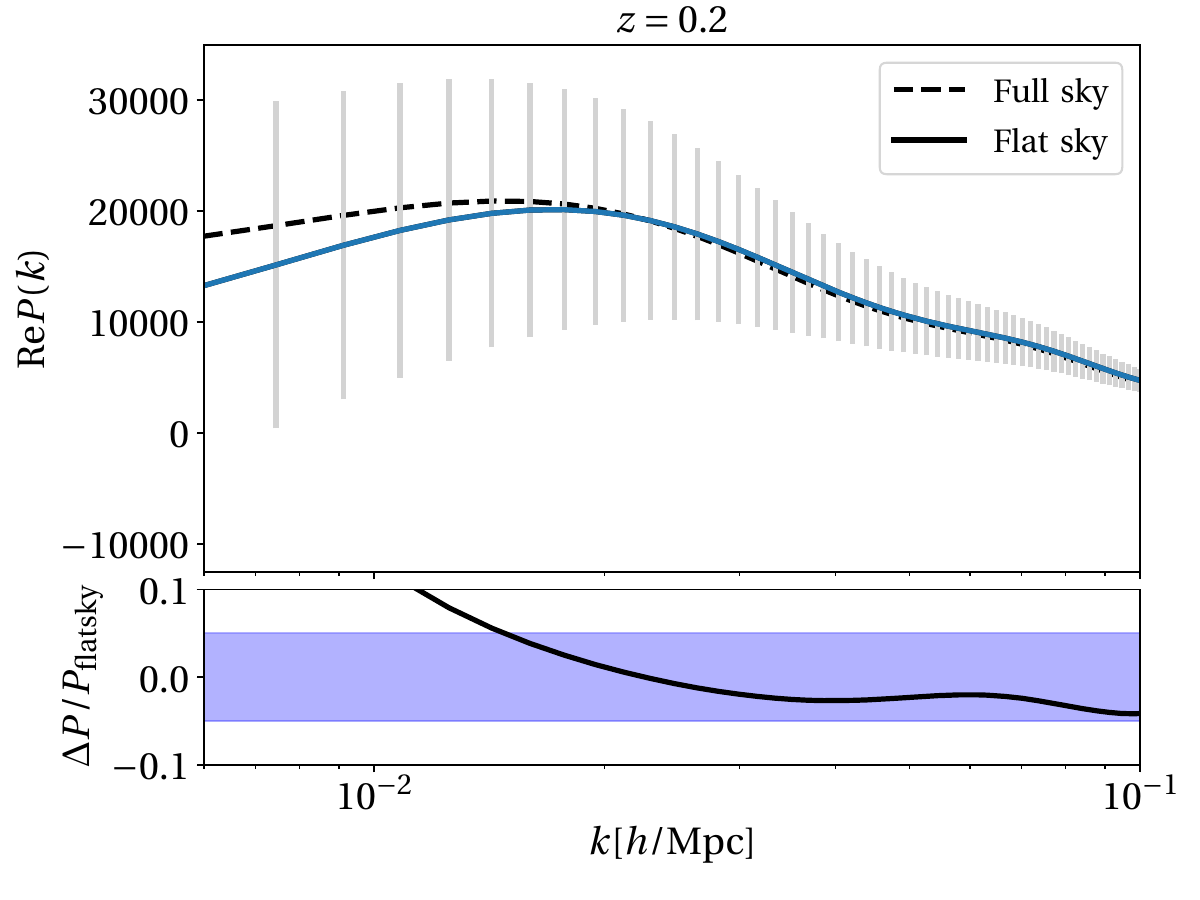}
    \includegraphics[width=0.32\linewidth]{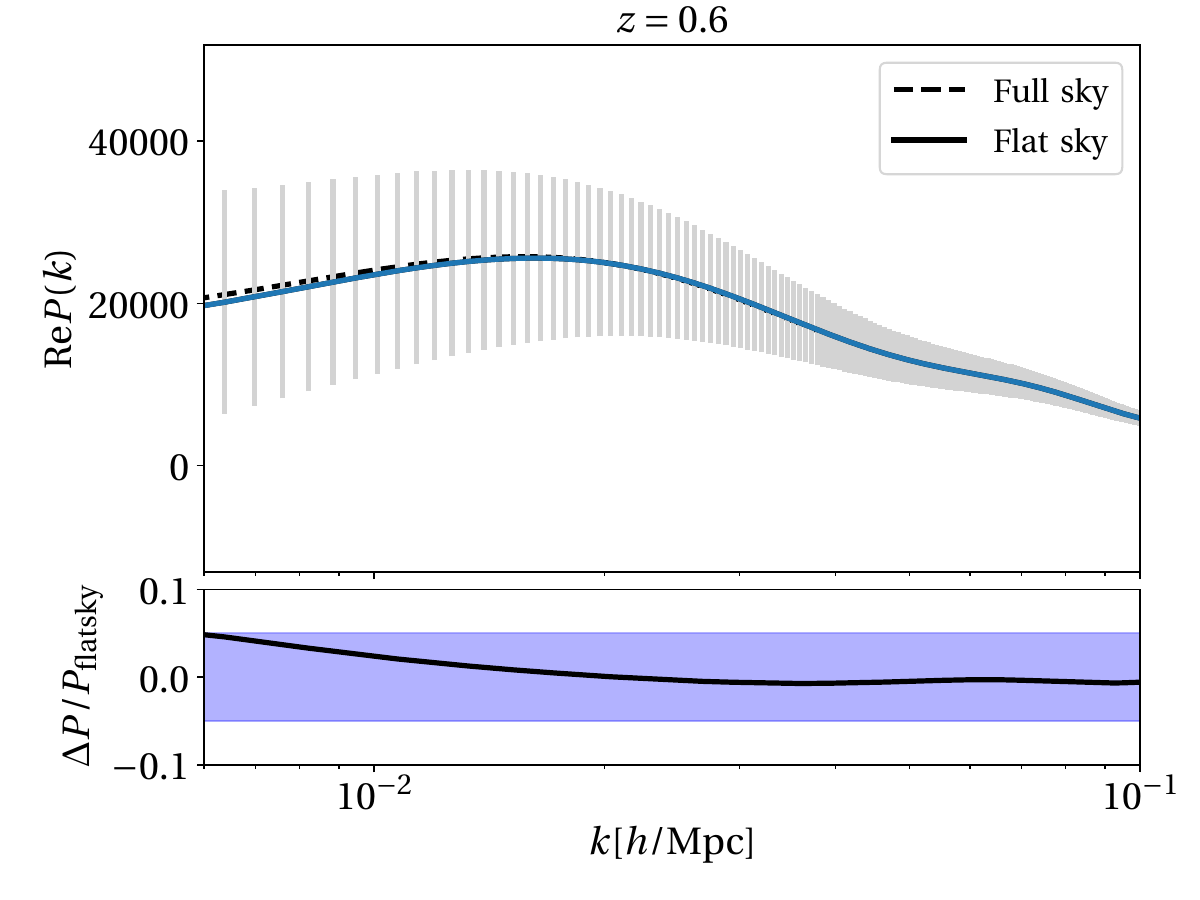}
    \includegraphics[width=0.32\linewidth]{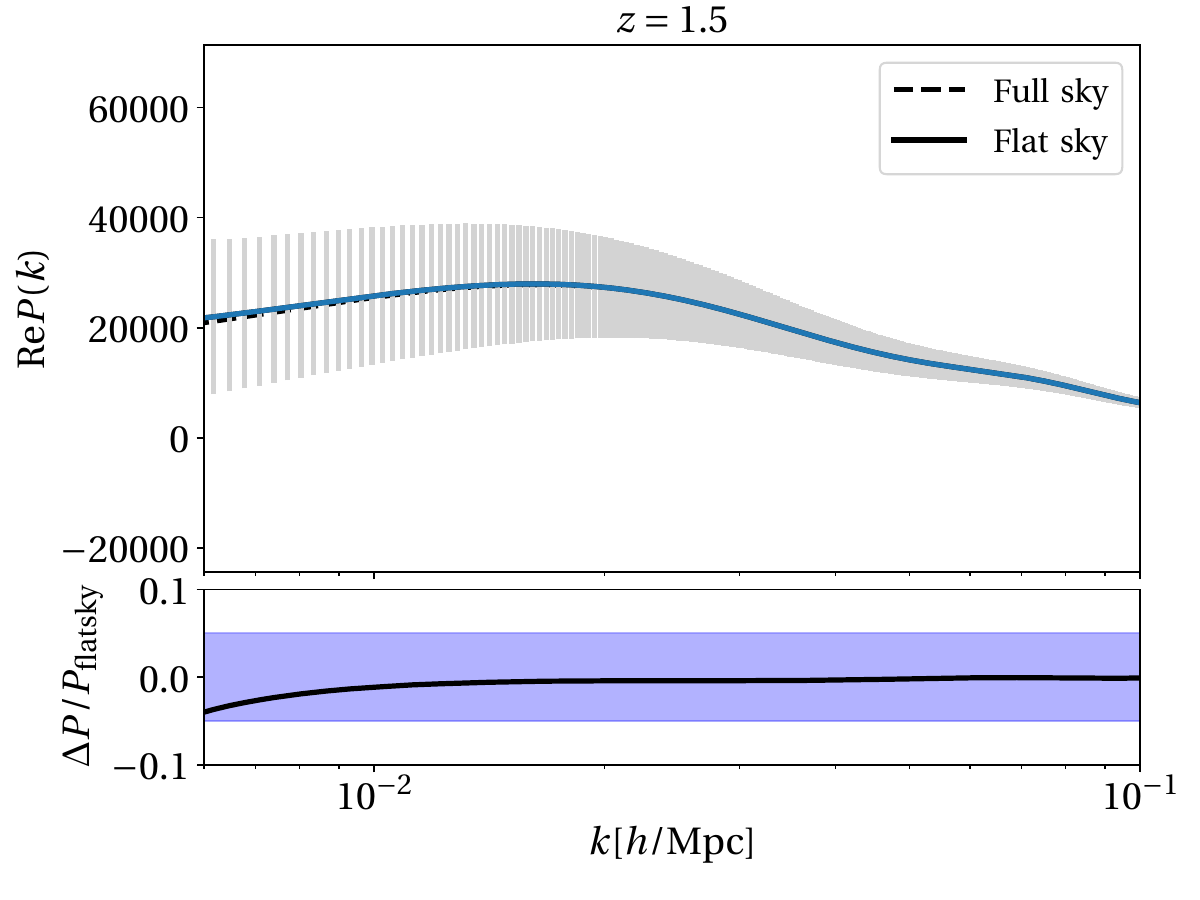}
    \caption{Comparison between the real part of the full-sky and flat-sky model in redshift space.
    Errors drawn from Equation~\eqref{eq:covkl} are also shown in grey.
    Residuals~$(P_\mathrm{full sky}-P_{\mathrm{flat sky}})/P_{\mathrm{flat sky}}$ are shown in the lower panels, where the shaded region highlights deviations~$\lesssim5\%$. }
    \label{fig:check_real}
\end{figure}

\begin{figure}
    \centering
    \includegraphics[width=0.32\linewidth]{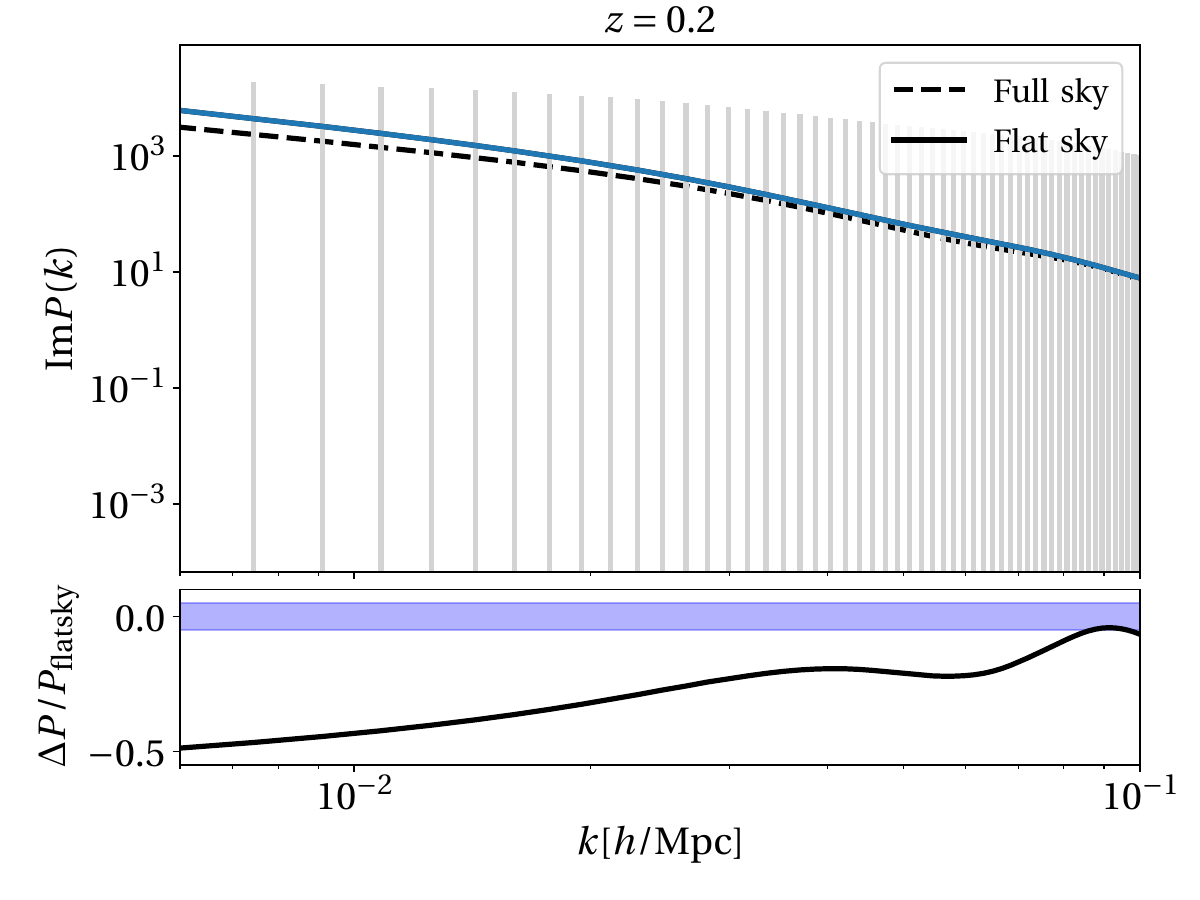}
    \includegraphics[width=0.32\linewidth]{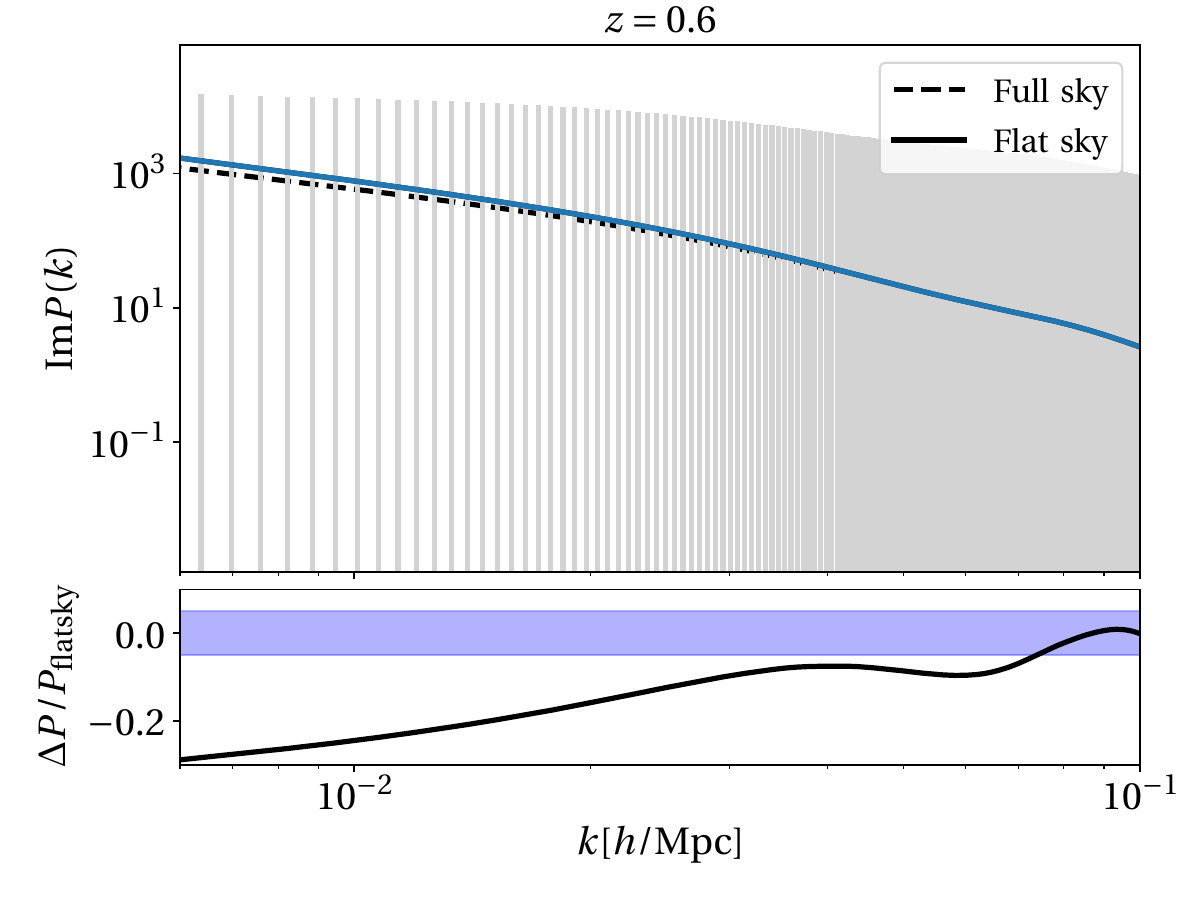}
    \includegraphics[width=0.32\linewidth]{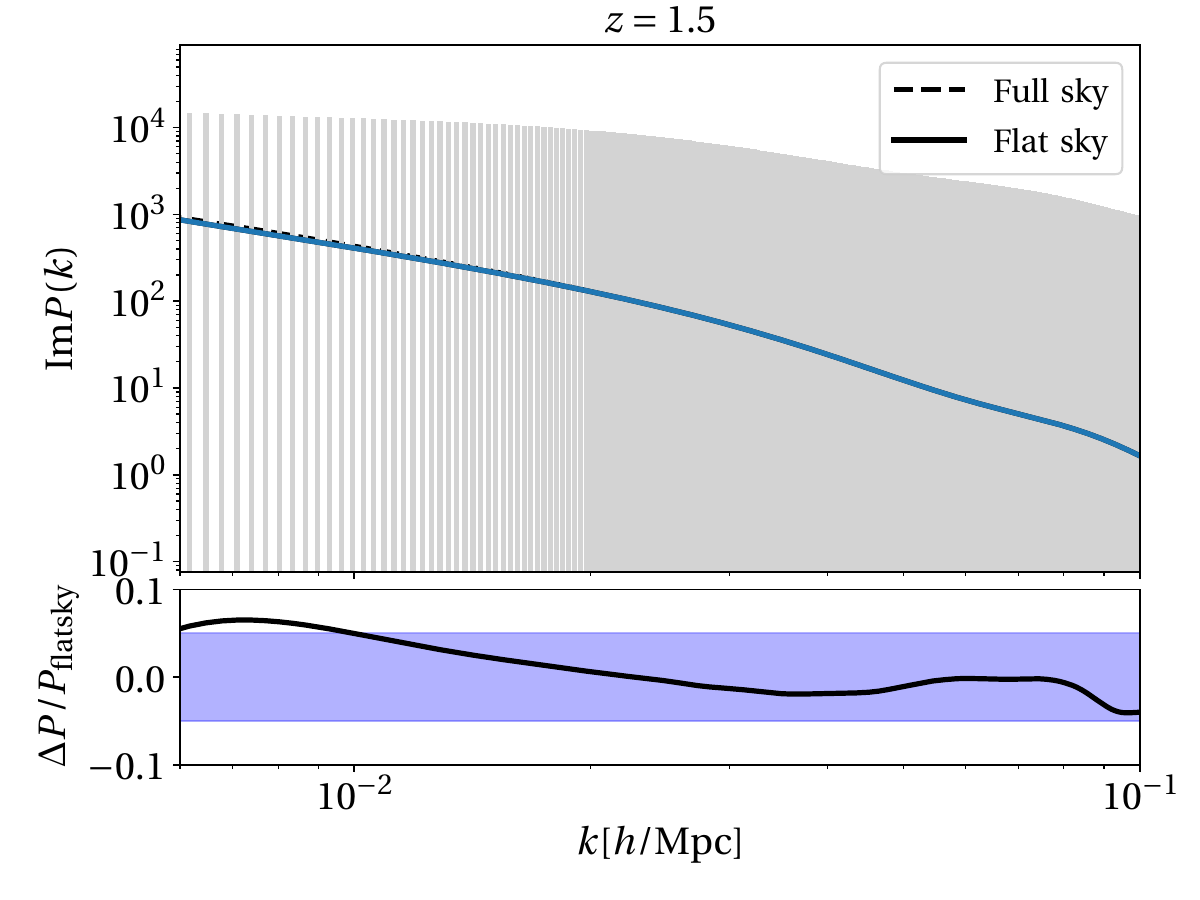}
    \caption{Same as Figure~\ref{fig:check_real}, but for the imaginary part of the power spectrum.}
    \label{fig:check_ima}
\end{figure}

To properly compare the two models, one needs to express everything in terms of the variables~$\{k_{\hat{n}},\ell/\bar{\chi}\}$. In these variables, the covariance of Equation~\eqref{eq:covMT} reads, neglecting shot noise terms,
\begin{equation}
\label{eq:covkl}
    \text{Cov}(k_{\hat{n}},\ell/\bar{\chi})=\bar{\chi}^2\frac{4\pi^2}{V\ell\Delta \ell\Delta k_{\hat{n}}}\left(P_{AB}^2+P_{AA}P_{BB}\right) \, .
\end{equation}
In Figure~\ref{fig:check_real} and Figure~\ref{fig:check_ima} we compare respectively the real and imaginary part of the models for different redshifts for $k_{\hat{n}}\sim0.003 \, h/{\rm Mpc}$, $\Delta k_{\hat{n}}=0.01 \, h/{\rm Mpc}$, $\Delta\ell=1$ as an example.
The redshift half-width is kept fixed at~$\Delta z=0.2$, and~$f_{\rm sky}\sim0.75$.
Errors taken from Equation~\eqref{eq:covkl} are also shown in grey.
Residuals~$(P_\mathrm{full sky}-P_{\mathrm{flat sky}})/P_{\mathrm{flat sky}}$ are shown in the lower panels, where the shaded region highlights deviations~$\lesssim5\%$.

For~$z\lesssim0.6$ the flat-sky model starts deviating considerably from the full-sky one at large scales, and this effect gets bigger at lower redshift, where the sky curvature becomes non-negligible for large scales.
While these kind of deviations seem to be well-within the cosmic variance limit, in this work we decided to focus on the redshift range~$z\gtrsim1$, where one can safely assume that the flat-sky approximation is very accurate.

\section{Relative importance of different unequal-time coefficients}\label{app:coeffs}
It can be instructive to investigate the relative importance of the various unequal-time corrections.

\begin{figure}
    \centering
    \includegraphics[width=1\linewidth]{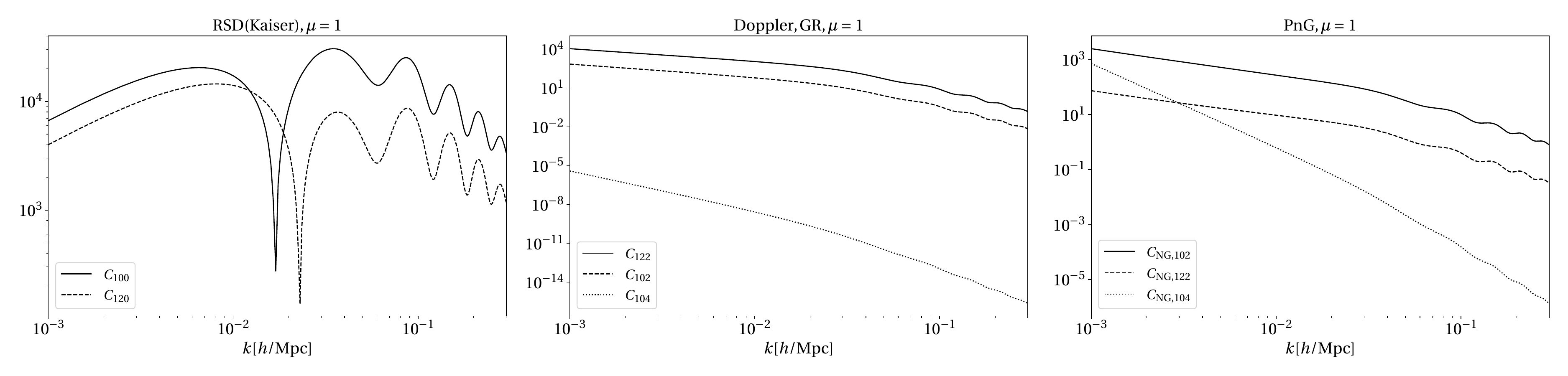}
    
    \caption{First order imaginary unequal-time corrections sourced by the RSD Kaiser term, for~$z=1.5$ for radial correlators.}
    \label{fig:imaradial}
\end{figure}

\begin{figure}
    \centering
    \includegraphics[width=0.7\linewidth]{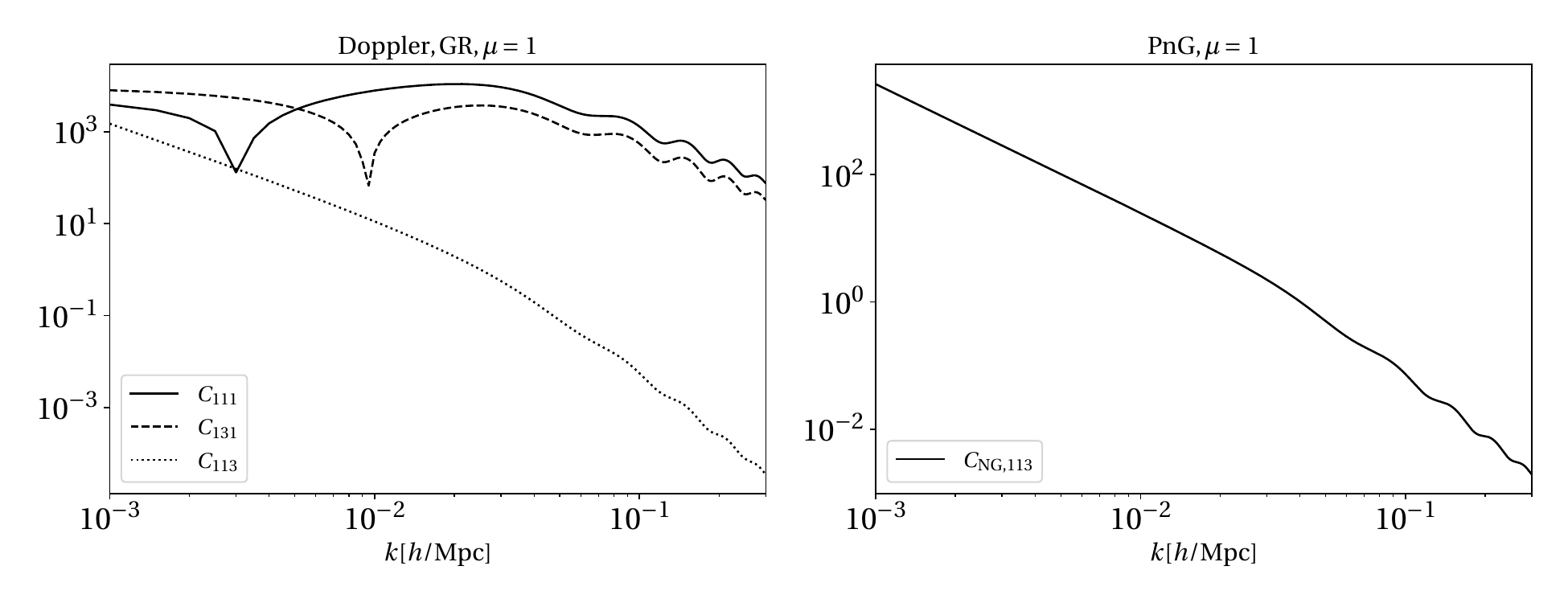}
    
    \caption{Same as Figure~\ref{fig:imaradial} but for real unequal-time corrections sourced by Doppler and GR terms.}
    \label{fig:real_radial}
\end{figure}

\begin{figure}
    \centering
    \includegraphics[width=1\linewidth]{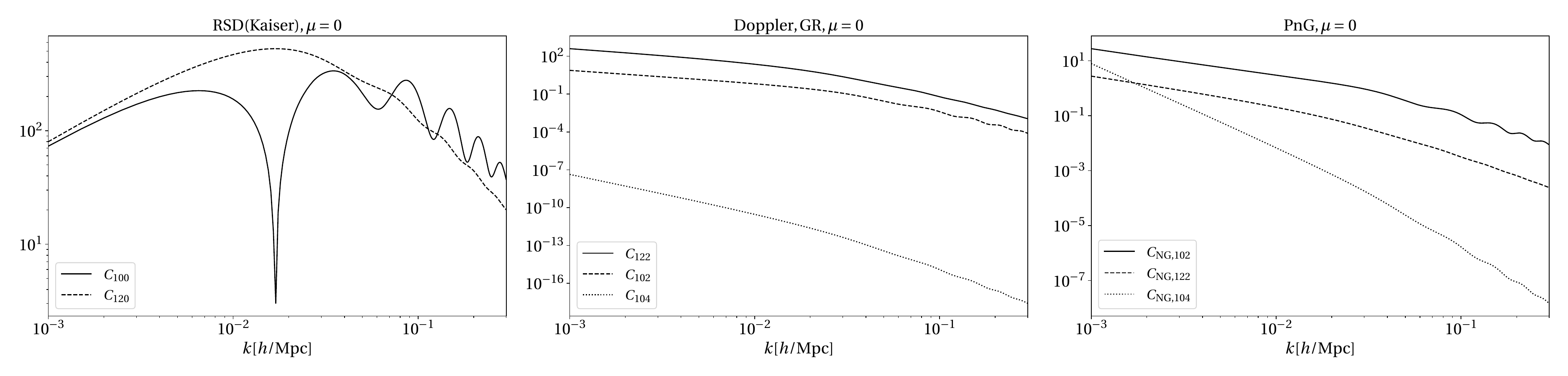}
    
    \caption{Same as Figure~\ref{fig:imaradial} but for transverse correlations.}
    \label{fig:imatransverse}
\end{figure}

\begin{figure}
    \centering
    \includegraphics[width=0.7\linewidth]{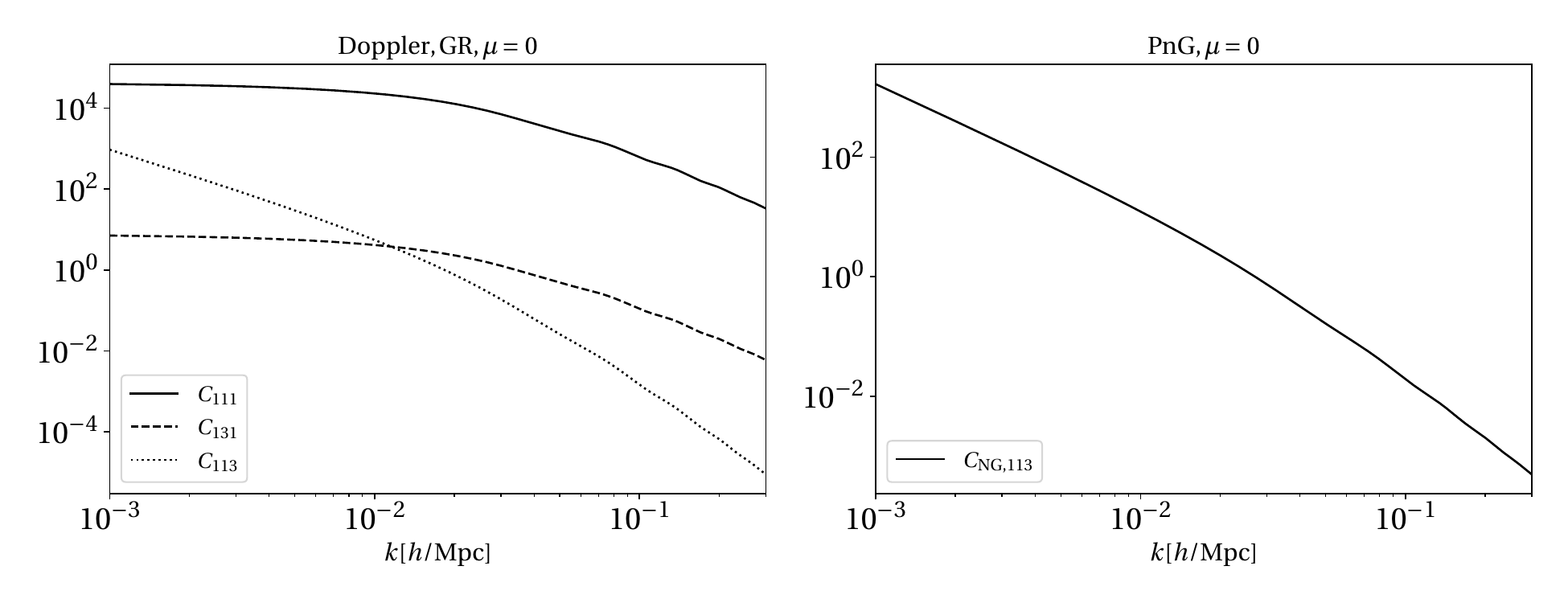}
    
    \caption{Same as Figure~\ref{fig:real_radial} but for transverse correlations.}
    \label{fig:realtransverse}
\end{figure}

Recall that since:
\begin{equation}
    \Delta_{\mathrm{UT}}\propto\frac{d}{dk_{\hat{n}}}\left[c_{1}(k,\mu,\bar{\chi}) \mathcal{P}_0(k) \right]\,,
\end{equation}
but $k$ and $\mu$ depend on $k_{\hat{n}}$ one gets
\begin{equation}
    \Delta_{\mathrm{UT}}\propto\frac{1}{k^{b+1}}\big[-b\mu^{a+1}+a\mu^{a-1}(1-\mu^2)\big]c_{1ab}(\bar{\chi}) \mathcal{P}_0(k)+  c_{1ab}(\bar{\chi})\frac{\mu^{a+1}}{k^b}\frac{d\mathcal{P}_0(k)}{dk}\bigg]\equiv C_{1ab}\,,
\end{equation}
where $c_{1ab}$ are given by Equations~\eqref{eq:c1ij_real} and Equations ~\eqref{eq:c1ij_real}.
Each $C_{1ab}$ gives the relative importance of various unequal-time corrections as a function of scales and orientation angle. They are plotted in Figure~\ref{fig:imaradial} and Figure~\ref{fig:real_radial} for $\mu=1$ and in Figure~\ref{fig:imatransverse} and Figure~\ref{fig:realtransverse} for $\mu=0$.
As one can see, the bulge of the unequal-time coefficients comes from $C_{100}$ which is imaginary and  $\propto b_B b'_A- b_A b'_B$ and for $\mu=1$. 
UT corrections arising from Doppler and GR terms, which contribute to the real part of unequal-time corrections, can be important at very large scales. In these case the dominant real UT correction is sourced by the coefficient $c_{111}$ and it is maximized for $\mu=0$.

\section{Impact of off-diagonal mode-mixing}
\label{app:offdiagonal}

Ref.~\cite{RVa} showed that the unequal-time angular power spectrum receives two distinct sets of corrections with respect to the equal-time case: the former set contains the genuine unequal-time corrections, coming from the Taylor expansion series in $\delta\chi$ of the Fourier-space kernels of the overdensity field; the latter set comes from the fact that, when~$\chi\neq\chi'$, the requirement that the transverse Fourier modes be equal introduces off-diagonal components to the Dirac delta~$\delta_D^{(2)}(\vec{\ell}-\vec{\ell}')$.
The full expression for the two-point correlator of the projected overdensity field is
\begin{equation}
    \VEV{\Hat{\delta}(\vec{\ell})\Hat{\delta}(\vec{\ell}\pr)} = (2\pi)^2 \delta_D^{(2)}(\vec{\ell}+\vec{\ell}\pr) \sum_{n=0}^{\infty} \frac{\left(\cevarr{\partial}_{\vec{\ell}\pr} \cdot \left(\vec{\ell}+\vec{\ell}\pr\right) \right)^n}{2^n n!} C^{(n)}_{\ell} \, ,
\end{equation}
with
\begin{equation}
    C^{(n)}_{\ell} = \int\frac{d\chi \, d\delta\chi}{\chi^2} \left(\frac{\delta\chi}{\chi}\right)^n W\left(\chi+\delta\chi/2\right) W'\left(\chi-\delta\chi/2\right) \int\frac{dk_n}{2\pi} e^{i\delta\chi k_n} \mathcal{P}(k_n\vhat{n},\vec{\ell}/\chi,\chi,\delta\chi) \, .
\end{equation}
Assuming Gaussian window functions
\begin{equation}
    W(\chi) = \frac{1}{\sqrt{2\pi}\sigma} e^{-(\chi-\chi_*)^2 / (2\sigma^2)} \, ,
\end{equation}
the coefficients can be written as
\begin{equation}
    C^{(n)}_{\ell} = \frac{1}{2\pi\sigma^2} \int\frac{d\chi}{\chi^{2+n}} D(\chi)^2 e^{-\frac{(\chi-\chi_*)^2}{\sigma^2}} \int\frac{dk_n}{2\pi} G^{(n,m)}(k_n,\chi) \mathcal{P}_0(k_n\vhat{n},\vec{\ell}/\chi) \, ,
\end{equation}
having defined the integral over the unequalness in time
\begin{equation}
    G^{(n,m)}(k_n,\chi) = \int d\delta\chi \, \delta\chi^n \left[ c_0(\chi) +\sum_{m=1}^{\infty} c_m(k_n,\ell,\chi) H^m \delta\chi^m \right] e^{i\delta\chi k_n -\frac{\delta\chi^2}{4\sigma^2}} \, .
\end{equation}
The~$n$ indices are related to the off-diagonal part of the delta function, while the~$m$ indices track the order of the Taylor expansion around the equal-time case.

In order to investigate the behavior of the different kinds of corrections in an analytical way, it is useful to work with a toy model for the power spectrum.
Assuming a power spectrum of the form
\begin{equation}
    \mathcal{P}_0(k) = A\left(\frac{k}{k_{\rm eq}}\right)^2 e^{-k^2/k_{\rm eq}^2} \, ,
\end{equation}
the results of the integration in~$k_n$ are
\begin{equation}
    I^{(n,m)}(\ell,\chi) = \int\frac{dk_n}{2\pi} G^{(n,m)}(k_n,\chi) \mathcal{P}_0(k_n\vhat{n},\vec{\ell}/\chi) \, ,
\end{equation}
which can be written in terms of the dimensionless variables
\begin{equation}
    \Xi^2 = 1 +(k_{\rm eq}\sigma)^2 \, , \qquad X = \frac{\ell/\chi}{k_{\rm eq}} \, .
\end{equation}

\begin{center}$n=0$\end{center}
\begin{align}
    & I^{(0,0)}(\ell,\chi) = \frac{A e^{-X^2}}{2X\Xi^3(k_{\rm eq}\sigma)^3} \Bigg\{ c_{000} X (\Xi^2-1)^2 (1+2 X^2\Xi^2) \nonumber \\
    &\qquad +c_{020} X (\Xi^2-1)^2 +c_{040} X (\Xi^2-1)^2 \left( 1-2X^2\Xi^2 +2 e^{X^2\Xi^2} \sqrt{\pi} X^3 \Xi^3 {\rm erfc}(X\Xi) \right) \nonumber \\
    &\qquad +c_{002} (H\sigma)^2 X \Xi^2 (\Xi^2-1) +2 c_{004} e^{X^2\Xi^2} (H\sigma)^4 \sqrt{\pi} {\rm erfc}(X\Xi) \nonumber \\
    &\qquad -2 c_{022} (H\sigma)^2 X \Xi^2 (\Xi^2-1) \left( e^{X^2\Xi^2}\sqrt{\pi} X\Xi {\rm erfc}(X\Xi) -1 \right) \Bigg\} \, , \\
    & I^{(0,1)}(\ell,\chi) = \frac{A e^{-X^2} (H \sigma)^2}{\Xi^3 (k_{\rm eq}\sigma)} \Bigg\{ c_{111}(1-\Xi^2) +2 c_{113} (H\sigma)^2 \Xi^2 \left( e^{X^2\Xi^2}\sqrt{\pi} X\Xi {\rm erfc}(X\Xi) -1 \right) \nonumber \\
    &\qquad +c_{131} (\Xi^2-1) \left( -1 +2X^2\Xi^2 -2 e^{X^2\Xi^2} \sqrt{\pi} X^3\Xi^3 {\rm erfc}(X\Xi) \right) \Bigg\} \, .
\end{align}

\begin{center}$n=1$\end{center}
\begin{align}
    & I^{(1,0)}(\ell,\chi) = \frac{A e^{-X^2} (H\sigma) \sigma}{\Xi^3 (k_{\rm eq}\sigma)} \Bigg\{ c_{011}(1-\Xi^2) +2 c_{013} (H\sigma)^2 \Xi^2 \left( e^{X^2\Xi^2} \sqrt{\pi} X\Xi {\rm erfc}(X\Xi) -1 \right) \nonumber \\
    &\qquad +c_{031} (\Xi^2-1) \left( -1 +2X^2\Xi^2 -2 e^{X^2\Xi^2} \sqrt{\pi} X^3\Xi^3 {\rm erfc}(X\Xi) \right) \Bigg\} \, , \\
    & I^{(1,1)}(\ell,\chi) = \frac{A e^{-X^2} (H\sigma) \sigma}{X\Xi^5 (k_{\rm eq}\sigma)^3} \Bigg\{ c_{100} X (\Xi^2-1)^2 (2\Xi^2-3) +2 c_{102} (H\sigma)^2 X \Xi^2 (\Xi^2-1) \nonumber \\
    &\qquad +2  c_{104} (H\sigma)^4 \Xi^4 \left[ -2X(\Xi^2-1) +e^{X^2\Xi^2} \sqrt{\pi} \Xi \left( 1+2X^2(\Xi^2-1) \right) {\rm erfc}(X\Xi) \right] \nonumber \\
    &\qquad +2c_{122} (H\sigma)^2 X\Xi^2 (\Xi^2-1) \left[ 1 +2X^2\Xi^2(\Xi^2-1) -e^{X^2\Xi^2} \sqrt{\pi} X\Xi^3 \left(1+2X^2(\Xi^2-1)\right) {\rm erfc}(X\Xi) \right] \Bigg\} \, .
\end{align}
Crucially, the diagonal~$I^{(0,1)}$ and the off-diagonal~$I^{(1,0)}$ are entirely sourced by the Doppler terms, which were not present in~\cite{RVa}.

\begin{center}$n=2$\end{center}
\begin{align}
    & I^{(2,0)}(\ell,\chi) = \frac{A e^{-X^2} \sigma^2}{\Xi^5 X (k_{\rm eq}\sigma)^3} \Bigg\{ c_{000} X (\Xi^2-1)^2 \left[ 3+2\Xi^2(X^2-1) \right] +2c_{002} (H\sigma)^2 X \Xi^2 (\Xi^2-1) \nonumber \\
    &\quad +2c_{004} \Xi^4(H\sigma)^4 \left[ -2X(\Xi^2-1) +e^{X^2\Xi^2} \sqrt{\pi} \Xi \left(1+2X^2(\Xi^2-1)\right) {\rm erfc}(X\Xi) \right] +c_{020} X (\Xi^2-1)^2 (3-2\Xi^2) \nonumber \\
    &\quad +c_{040} X (\Xi^2-1)^2 \left[ 3-2\Xi^2 \left( 1+X^2+2X^4\Xi^2(\Xi^2-1) \right) +2e^{X^2\Xi^2} \sqrt{\pi} X^3 \Xi^5 \left(1+2X^2(\Xi^2-1)\right) {\rm erfc}(X\Xi) \right] \nonumber \\
    &\quad +2c_{022} (H\sigma)^2 X\Xi^2 (\Xi^2-1) \left[ 1+2X^2\Xi^2(\Xi^2-1) -e^{X^2\Xi^2} \sqrt{\pi} X\Xi^3 \left(1+2X^2(\Xi^2-1)\right) {\rm erfc}(X\Xi) \right] \Bigg\} \, , \\
    & I^{(2,1)}(\ell,\chi) = -\frac{2A e^{-X^2} (H\sigma)^2 \sigma^2}{\Xi^5 (k_{\rm eq}\sigma)} \Bigg\{ 3c_{111}(\Xi^2-1) \nonumber \\
    &\quad -2c_{113} (H\sigma)^2 \Xi^2 \left[ -1+2\Xi^2(X^2-1) -2X^2\Xi^4 +e^{X^2\Xi^2} \sqrt{\pi} X\Xi^3 \left( 3+2X^2(\Xi^2-1) \right) {\rm erfc}(X\Xi) \right] \nonumber \\
    &\quad +c_{131} (\Xi^2-1) \left[ 3-2X^2\Xi^2 \left( 1+2\Xi^2(1+X^2(\Xi^2-1)) \right) +2e^{X^2\Xi^2} \sqrt{\pi} X^3\Xi^5 \left(3+2X^2(\Xi^2-1)\right) {\rm erfc}(X\Xi) \right] \Bigg\} \, .
\end{align}

The coefficients are
\begin{equation}
\begin{split}
    c_0(k_n,\ell/\chi,\chi) =& \left[ b_A +f\mu^2 +\frac{\mathcal{A}_A}{k^2} -\frac{if\mu\alpha_A}{k\chi} \right] \left[ b_B +f\mu^2 +\frac{\mathcal{A}_B}{k^2} +\frac{if\mu\alpha_B}{k\chi} \right] \\
    =& c_{000} +c_{020} \mu^2 +c_{040} \mu^4 +c_{002} \left(\frac{H}{k}\right)^2 +c_{022} \left(\frac{H}{k}\right)^2 \mu^2 +c_{004} \left(\frac{H}{k}\right)^4 \\
    & +i c_{011} \mu \left(\frac{H}{k}\right) +i c_{031} \mu^3 \left(\frac{H}{k}\right) +i c_{013} \mu \left(\frac{H}{k}\right)^3 \, ,
\end{split}
\end{equation}
with
\begin{subequations}
\begin{align}
    c_{000} &= b_Ab_B \, , \\
    c_{020} &= f(b_A+b_B) \, , \\
    c_{040} &= f^2 \, , \\
    c_{002} &= \frac{\mathcal{A}_Bb_A+\mathcal{A}_Ab_B}{H^2} \, , \\
    c_{022} &= \frac{f\left(f\alpha_A\alpha_B+(\mathcal{A}_A+\mathcal{A}_B)\chi^2\right)}{(H\chi)^2} \, , \\
    c_{004} &= \frac{\mathcal{A}_A\mathcal{A}_B}{H^4} \, , \\
    c_{011} &= \frac{f(b_A\alpha_B-b_B\alpha_A)}{H\chi} \, , \\
    c_{031} &= \frac{-f^2(\alpha_A-\alpha_B)}{H\chi} \, , \\
    c_{013} &= \frac{f(\mathcal{A}_A\alpha_B-\mathcal{A}_B\alpha_A)}{H^3 \chi} \, ,
\end{align}
\end{subequations}
and
\begin{equation}
\begin{split}
    c_1(k_n,\ell/\chi,\chi) =& c_{100} +c_{120} \mu^2 +c_{102} \left(\frac{H}{k}\right)^2 +c_{122} \mu^2 \left(\frac{H}{k}\right)^2 +c_{104} \left(\frac{H}{k}\right)^4 \\
    & +i c_{111} \mu\left(\frac{H}{k}\right) +i c_{131} \mu^3 \left(\frac{H}{k}\right) +i c_{113} \mu \left(\frac{H}{k}\right)^3 \, ,
\end{split}
\end{equation}
where, as in the main text,
\begin{subequations}
\begin{align}
    c_{100} &= \frac{1}{2}  b_A b_B \left( \frac{b_A'}{b_A}-\frac{b_B'}{b_B} \right) \, , \\
   c_{120} &= \frac{f}{2}\left[ b_A' -b_B' -\frac{f'}{f}\left(b_A-b_B\right) \right] \, , \\
    c_{102} &= \frac{1}{2}\left[ \left( b_A' \frac{\mathcal{A}_B}{H^2} -b_A \frac{\mathcal{A}_B'}{H^2} \right) -\left( b_B' \frac{\mathcal{A}_A}{H^2} -b_B \frac{\mathcal{A}_A'}{H^2} \right) \right] \, , \\
    c_{122} &= \frac{f}{2}\left[ \frac{\mathcal{A}_A'}{H^2}-\frac{\mathcal{A}_B'}{H^2} -\frac{f'}{f}\left(\frac{\mathcal{A}_A}{H^2}-\frac{\mathcal{A}_B}{H^2}\right) +\frac{f \alpha_A \alpha_B}{(H\chi)^2} \left( \frac{\alpha_A'}{\alpha_A}-\frac{\alpha_B'}{\alpha_B} \right) \right] \, , \\
    c_{104} &= \frac{1}{2} \frac{\mathcal{A}_A}{H^2} \frac{\mathcal{A}_B}{H^2} \left( \frac{\mathcal{A}_A'}{\mathcal{A}_A} -\frac{\mathcal{A}_B'}{\mathcal{A}_B} \right) \, , \\
    c_{111} &= \frac{f}{2 H\chi} \left[ b_A'\alpha_B+b_B'\alpha_A -b_A\alpha_B' -b_B\alpha_A' +\left(b_A\alpha_B+b_B\alpha_A\right) \left(\frac{1}{H\chi}-\frac{f'}{f}\right) \right] \, , \\
    c_{131} &= \frac{f^2}{ 2H\chi} \left( \frac{\alpha_A+\alpha_B}{H\chi} -\alpha_A' -\alpha_B' \right) \, , \\
    c_{113} &= \frac{f}{ 2H\chi} \left[ \left(\alpha_A\frac{\mathcal{A}_B}{H^2}+\alpha_B\frac{\mathcal{A}_A}{H^2} \right)\left(\frac{1}{H\chi}-\frac{f'}{f}\right) +\alpha_A\frac{\mathcal{A}_B'}{H^2} +\alpha_B\frac{\mathcal{A}_A'}{H^2} -\alpha_A'\frac{\mathcal{A}_B}{H^2} -\alpha_B'\frac{\mathcal{A}_A}{H^2} \right] \, .
\end{align}
\end{subequations}

\begin{figure}
    \centering
    \includegraphics[width=0.8\linewidth]{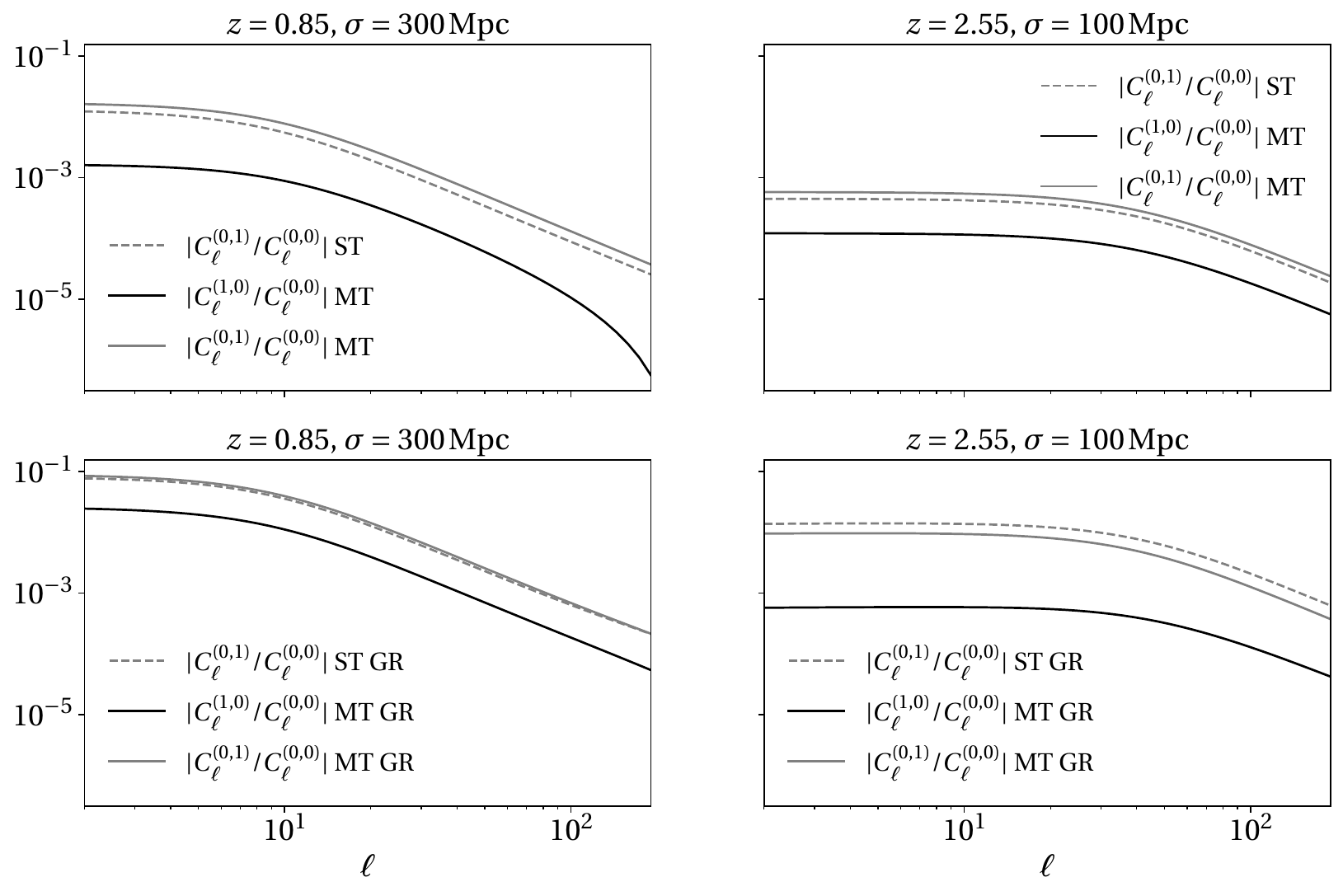}
    \caption{Ratio of the first-order corrections in~$\delta\chi$ to the equal-time, diagonal angular power spectrum~$C_{\ell}^{(0,0)}$.}
    \label{fig:offdiagonal}
\end{figure}

Figure~\ref{fig:offdiagonal} shows the relative importance of the first-order corrections in~$\delta\chi$, arising from either mode-mixing due to off-diagonal terms in the Dirac delta~$n=1$, or unequal-time expansion of the overdensity field kernels~$m=1$.
We confirm our expectation that the set of correction arising from the off-diagonal part is negligible with respect to the genuinely unequal-time Taylor expansion, even when including Doppler terms and GR corrections.

\section{Single tracer}
\label{app:single}

If we consider only the Kaiser RSD kernel, UT corrections vanish at first order in the single-tracer scenario.
This is not the case when we also include Doppler terms; the first-order (for the Gaussian case) UT correction for a single tracer is purely real and reads
\begin{equation}
    c_{1,\rm ST}(k,\mu,\bar{z}) = \frac{i \mu f b \alpha}{k\chi} \left[ \frac{b'}{b} -\frac{\alpha'}{\alpha} +\left(\frac{1}{H\chi}-\frac{f'}{f}\right) \right] +\frac{i \mu^3 f^2 \alpha}{k\chi} \left( \frac{1}{H\chi} -\frac{\alpha'}{\alpha} \right) +\frac{i \mu f \alpha \mathcal{A}}{k^3\chi} \left[ \left(\frac{1}{H\chi}-\frac{f'}{f}\right) -\frac{\alpha'}{\alpha} +\frac{\mathcal{A}'}{\mathcal{A}} \right] \, ,
\end{equation}
while the non-Gaussian part is
\begin{equation}
    c_{1,\rm ST, NG}(k,\mu,\bar{z})=\frac{3\fnl\Omega_{m,0} H_0^2}{2 T(k) D_{\rm md}(z)}\frac{f \alpha b_{\phi}}{H^3\chi}\left[\frac{b_{\phi}'}{b_{\phi}}+\frac{f}{1+z} +\left(\frac{1}{H\chi}-\frac{f'}{f}\right)-\frac{\alpha'}{\alpha}\right] \, .
\end{equation}
The SNR of the UT correction in the single-tracer case can be calculated in the same fashion as in Equation~\eqref{eq:SNR}. However one can show that for Stage IV-like surveys the ${\rm SNR} \ll 1$ making such corrections completely negligible.

\end{document}